\documentclass[reprint,superscriptaddress,amsmath,amssymb,twocolumn]{revtex4-2}

\usepackage{graphicx}
\usepackage{dcolumn}
\usepackage{bm}
\usepackage{microtype}
\usepackage{array}
\usepackage[usenames,dvipsnames]{xcolor}
\definecolor{nblue}{rgb}{0.0, 0.0, 1.0}
\definecolor{blue}{rgb}{0.0, 0.0, 1.0}
\definecolor{magenta}{rgb}{0.79, 0.08, 0.48}
\usepackage[colorlinks,linkcolor=nblue,urlcolor=magenta,citecolor=magenta,plainpages=false,pdfpagelabels,breaklinks]{hyperref}

\newcommand{\beq}{\begin{equation}}
\newcommand{\eeq}{\end{equation}}
\newcommand{\bea}{\begin{eqnarray}}
\newcommand{\eea}{\end{eqnarray}}

\newcommand{\bk} { \bm{k} }
\newcommand{\bd} { \bm{d} }
\newcommand{\bq} { \bm{q} }

\newcommand{\tcr}[1]{{\color[rgb]{0,0,0}{#1}}}

\renewcommand{\figurename}{Fig.}
\renewcommand{\tablename}{Table}
\makeatletter\renewcommand{\fnum@figure}[1]{\textbf{\figurename~\thefigure\,\textbar\,}}\makeatother
\makeatletter\renewcommand{\fnum@table}[1]{\tablename~\thetable\,\textbar\,}\makeatother

\begin{document}



\title{Spin-triplet superconductivity in Weyl nodal-line semimetals}






\author{Tian\ Shang}\thanks{These authors contributed equally}
\affiliation{Key Laboratory of Polar Materials and Devices (MOE), School of Physics and Electronic Science, East China Normal University, Shanghai 200241, China}
\author{Sudeep\ K.\ Ghosh}\thanks{These authors contributed equally}
\affiliation{School of Physical Sciences, University of Kent, Canterbury CT2 7NH, United Kingdom}

\author{Michael\ Smidman}\thanks{These authors contributed equally}
\affiliation{Center for Correlated Matter and Department of Physics, Zhejiang University, Hangzhou 310058, China}
\author{Dariusz Jakub Gawryluk} 
\affiliation{Laboratory for Multiscale Materials Experiments, Paul Scherrer Institut, CH-5232 Villigen PSI, Switzerland}
\author{Christopher Baines}
\affiliation{Laboratory for Muon-Spin Spectroscopy, Paul Scherrer Institut, CH-5232 Villigen PSI, Switzerland}
\author{An Wang}
\affiliation{Center for Correlated Matter and Department of Physics, Zhejiang University, Hangzhou 310058, China}
\author{Wu Xie}
\affiliation{Center for Correlated Matter and Department of Physics, Zhejiang University, Hangzhou 310058, China}
\author{Ye Chen}
\affiliation{Center for Correlated Matter and Department of Physics, Zhejiang University, Hangzhou 310058, China}
\author{Mukkattu O. Ajeesh}
\affiliation{Max Planck Institute for Chemical Physics of Solids, N\"{o}thnitzer Str.40, 01187, Dresden, Germany}
\author{Michael Nicklas}
\affiliation{Max Planck Institute for Chemical Physics of Solids, N\"{o}thnitzer Str.40, 01187, Dresden, Germany}
\author{Ekaterina Pomjakushina}
\affiliation{Laboratory for Multiscale Materials Experiments, Paul Scherrer Institut, CH-5232 Villigen PSI, Switzerland}
\author{Marisa Medarde}
\affiliation{Laboratory for Multiscale Materials Experiments, Paul Scherrer Institut, CH-5232 Villigen PSI, Switzerland}
\author{Ming Shi}
\affiliation{Swiss Light Source, Paul Scherrer Institut, CH-5232 Villigen PSI, Switzerland}
%
\author{James F. Annett}
\affiliation{H. H. Wills Physics Laboratory, University of Bristol, Tyndall Avenue, Bristol BS8 1TL, United Kingdom}
\author{Huiqiu\ Yuan}\email[Electronic address: ]{hqyuan@zju.edu.cn}
\affiliation{Center for Correlated Matter and Department of Physics, Zhejiang University, Hangzhou 310058, China}
\author{Jorge\ Quintanilla}\email[Electronic address: ]{j.quintanilla@kent.ac.uk}
\affiliation{School of Physical Sciences, University of Kent, Canterbury CT2 7NH, United Kingdom}
\author{Toni Shiroka}\email[Electronic address: ]{tshiroka@phys.ethz.ch}
\affiliation{Laboratory for Muon-Spin Spectroscopy, Paul Scherrer Institut, CH-5232 Villigen PSI, Switzerland}
\affiliation{Laboratorium f\"ur Festk\"orperphysik, ETH Z\"urich, CH-8093 Z\"urich, Switzerland}
%
%


\date{\today}


\begin{abstract}
Topological semimetals are three dimensional materials with symmetry-protected
massless bulk excitations. As a special case, Weyl nodal-line semimetals are realized in materials either having no inversion or broken time-reversal symmetry and feature bulk nodal lines. The 111-family of materials, LaNiSi, LaPtSi and LaPtGe (all lacking inversion symmetry), belong to this class.
%
Here, by combining muon-spin rotation and relaxation- with thermodynamic 
measurements, we find that these materials exhibit a fully-gapped superconducting 
ground state, while spontaneously breaking time-reversal symmetry at 
the superconducting transition. Since time-reversal symmetry is essential for protecting the normal-state 
topology, its breaking upon entering the superconducting state should 
remarkably result in a topological phase transition. 
%
By developing a minimal model for the normal-state band structure and assuming 
a purely spin-triplet pairing, we show that the superconducting properties 
across the family can be described accurately. 
%
%
Our results demonstrate that the 111-family reported here provides an 
ideal test-bed for investigating the rich interplay between the exotic 
properties of Weyl nodal-line fermions and unconventional superconductivity.
\end{abstract}
\maketitle

Topological materials are at the forefront of current condensed matter
and material science research due to their great potential for applications.
Among the defining characteristics of topological materials is their 
symmetry-protected metallic surface state, arising from a nontrivial
bulk topology. Recently, the experimental observation of many topological
semimetals has shifted the research focus towards this subclass of
topological materials~\cite{Armitage2018,Lv2021}. Contrary to Dirac- or
Weyl semimetals, which have point-type band crossings, in nodal-line
semimetals band crossings occur in the form of lines or rings along
special $\bk$-directions of the Brillouin zone. In this case, near the
nodes, the low-energy excitations are nodal-line fermions with rather
exotic properties~\cite{Armitage2018,Lv2021}. Weyl nodal-line semimetals
can be realized in systems lacking inversion symmetry or with broken
time-reversal symmetry (TRS), provided the nodal lines
are protected by additional symmetries. 
Recently, the isostructural noncentrosymmetric 111-type materials LaNiSi,
LaPtSi, and LaPtGe have been predicted to be Weyl nodal-line semimetals, 
protected by nonsymmorphic glide planes~\cite{Zhang2020}. In addition,
at low temperatures, all of them become superconductors~\cite{Lee1994,Kneidinger2013,Evers1984}.

The breaking of additional symmetries in the superconducting state,
besides the global gauge symmetry of the wave function, is a key
characteristic of unconventional superconductors~\cite{Annett1990,Sigrist1991}.
The combination of intriguing fundamental physics with far-reaching
potential for applications has made unconventional superconductors one
of the most investigated classes of materials. Broken time reversal
symmetry in the superconducting state, one of the typical indications
of unconventional super\-conduct\-ivity (SC), is manifested by the spontaneous appearance
of magnetic fields below the superconducting transition temperature
$T_c$~\cite{Ghosh2020a}. Recently, by using the muon-spin relaxation technique, several noncentrosymmetric superconductors (NCSCs)
have been found to break TRS in their superconducting state.  
Otherwise they appear to exhibit the conventional properties of
standard phonon-mediated superconductors~\cite{Ghosh2020a,Hillier2009,Barker2015,Singh2014,Shang2018b,Shang2020a}.
In NCSCs, singlet-triplet admixed pairings can be induced by antisymmetric
spin-orbit coupling (ASOC), however, ASOC itself cannot break
TRS~\cite{smidman2017,Bauer2012}. Noncentrosymmetric superconductors also
provide a fertile ground also for topological superconductivity, with
potential applications to topological quantum computing~\cite{Sato2017,Qi2011,Kallin2016}.

According to electronic band-structure calculations, the ASOC strength
increases progressively from LaNiSi to LaPtSi to LaPtGe~\cite{Zhang2020}.
Hence, the 111-family of materials is a prime 
candidate for investigating the relationship between ASOC and 
unconventional SC with TRS-breaking, here made even more 
interesting by the interplay with the exotic nodal-line fermions. 
\tcr{Recent muon-spin relaxation and rotation ($\mu$SR) studies 
on LaNiSi and LaPtSi reported an enhanced muon-spin relaxation 
at low temperatures, seemingly an indication of TRS breaking. 
However, their unusual temperature dependence (here resembling a 
Curie-Weiss behavior), the lack of any distinct features near 
$T_c$~\cite{Sajilesh2020}, and the absence of an additional muon-spin 
relaxation in LaPtGe (below its $T_c$)~\cite{Sajilesh2018}, 
all
suggest that TRS is preserved in the superconducting 
state of these 111 materials. 
We recall that, in the past, inconsistent $\mu$SR results have been 
reported in UPt$_3$~\cite{Luke1993,Dalmas1995}, whose TRS breaking 
could be independently proved by optical Kerr effect only a decade later~\cite{Schemm2014}. 
Clearly, it is highly desirable to investigate also the 111 materials 
with other techniques such as the Kerr effect, in 
order to confirm their TRS breaking.} 
Here, by combining extensive and thorough $\mu$SR 
measurements with detailed theoretical analysis, we show that, 
contrary to previous reports, all the 
above 111-type materials spontaneously break TRS at the superconducting
transition and exhibit a fully-gapped pure spin-triplet pairing.\\

\begin{figure}[!t]
	\begin{center}
		\includegraphics[width=0.48\textwidth]{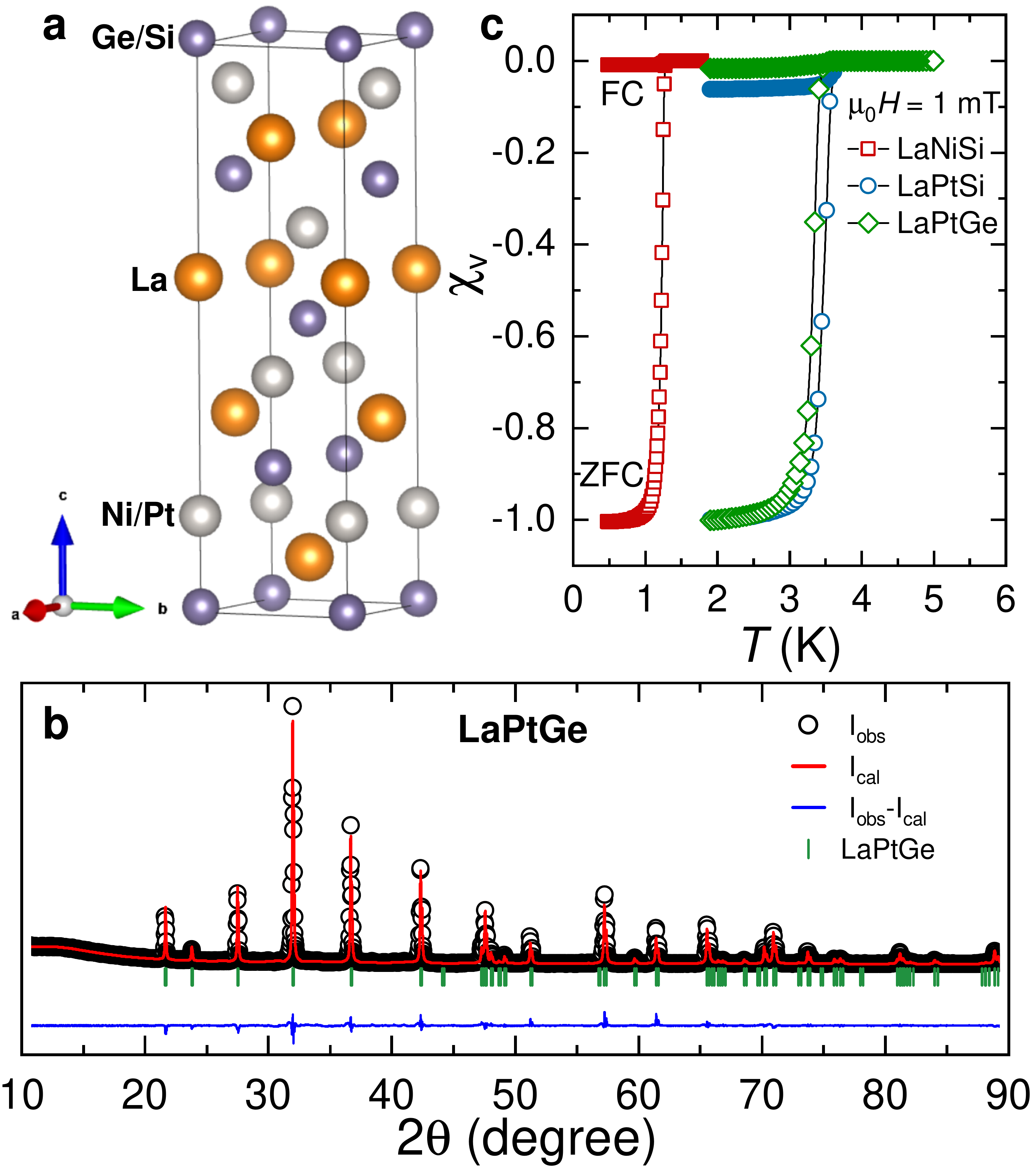}
	\end{center}
	\caption{\label{fig:bulkSC}\textbf{Crystal structure and bulk superconductivity}. \textbf{a}, Crystal structure of LaNiSi, LaPtSi, and LaPtGe. 
	\tcr{\textbf{b}, Room-temperature XRD pattern and Rietveld refinements for LaPtGe.  The black circles and the solid-red line represent the experimental pattern and the Rietveld refinement	profile, respectively. The blue line at the bottom shows the residuals, i.e., the difference between calculated and experimental
    data. The vertical bars mark the calculated Bragg-peak positions for LaPtGe. The  Rietveld refinements of LaNiSi and LaPtSi are shown in the Suppl.\ Fig.~S1.}
	\textbf{c}, Temperature dependence of the magnetic susceptibility. The zero field-cooled (ZFC) and field-cooled (FC) magnetic susceptibility were measured in a field of $\mu_{0}H$ = 1\,mT. The well distinct 
    ZFC- and FC curves are consistent with type-II SC, 
    as confirmed also by $\mu$SR measurements.}
\end{figure}

\begin{figure}[!t]
	\begin{center}
		\includegraphics[width=0.48\textwidth]{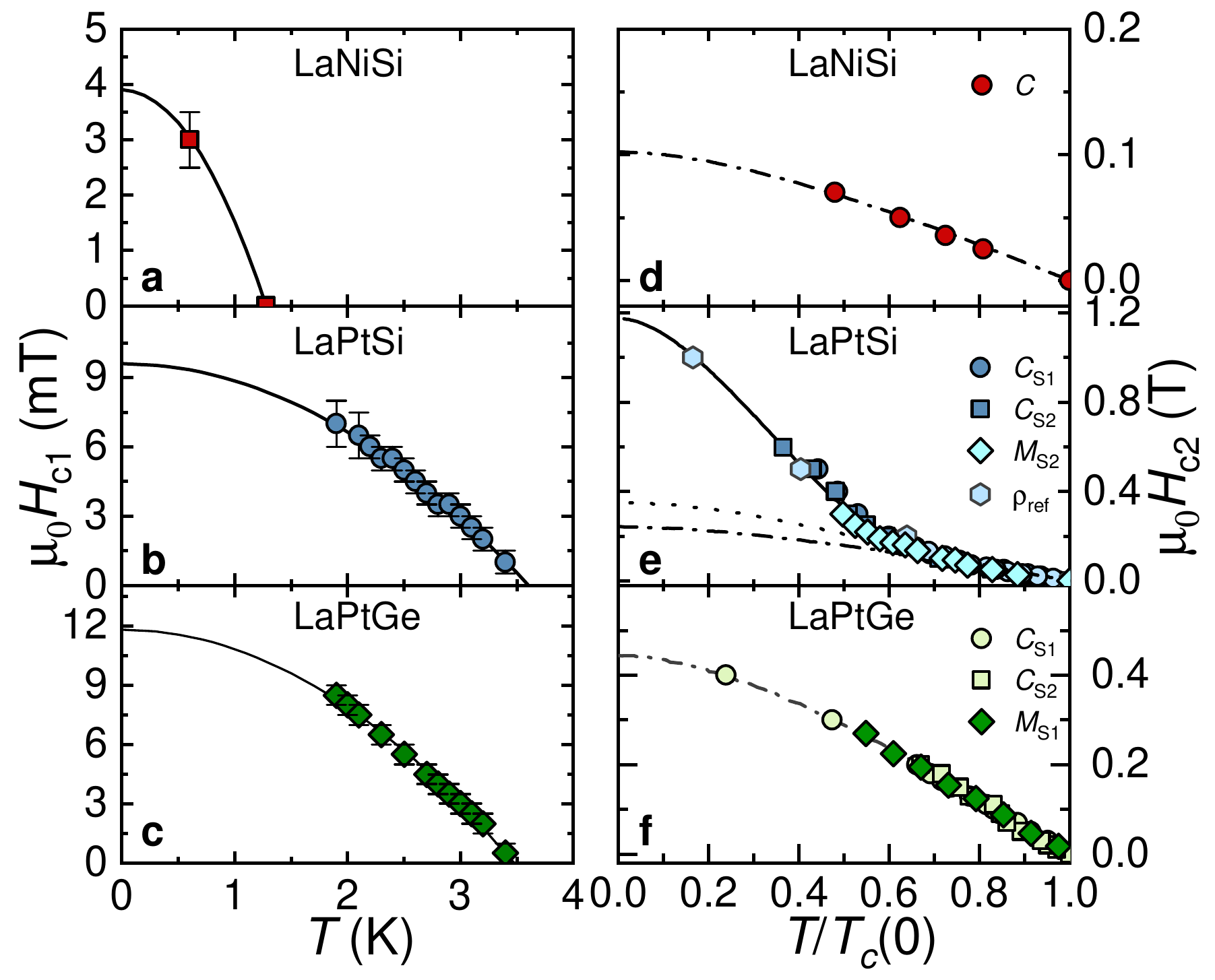}
	\end{center}
	\caption{\label{fig:field} \tcr{
		\textbf{Lower- and upper critical fields}. \textbf{a, b, c,} The lower critical fields $H_{c1}$ as a function of temperature for LaNiSi (\textbf{a}), LaPtSi (\textbf{b}), and LaPtGe (\textbf{c}). 
		For each temperature, $H_{c1}$ was determined as the value where the field-dependent magnetization $M(H)$ starts to deviate from linearity (see Fig. S3 in Suppl.\ Mat.). Solid lines are fits to 
		$\mu_{0}H_{c1}(T) =\mu_{0}H_{c1}(0)[1-(T/T_{c})^2]$.
		\textbf{d, e, f,} Upper critical fields $H_{c2}$ versus the reduced temperature $T/T_c$ for LaNiSi (\textbf{d}), LaPtSi (\textbf{e}), and LaPtGe (\textbf{f}). The superconducting transition temperatures $T_c$ were determined from  heat-capacity- 
		$C(T)$ and magnetization measurements $M(H)$ (see details in Figs. S4--S6 in Suppl.\ Mat.). For LaPtSi and LaPtGe, two different sample batches (denoted as S1 and S2) 
		were measured.  
		The dash-dotted-, dashed-, and solid lines are fits using WHH-, GL-, and two-band models, respectively. 
		The $H_\mathrm{c2}$ values determined from $\rho(T,H)$ from Ref.~\onlinecite{Kneidinger2013} are also plotted in \textbf{e} for LaPtSi. The error bars of $H_{c1}$ are the field-step values used during the $M(H)$ measurements.}
	}
\end{figure}

\begin{figure*}[htb]
\begin{center}
 \includegraphics[width=0.8\textwidth]{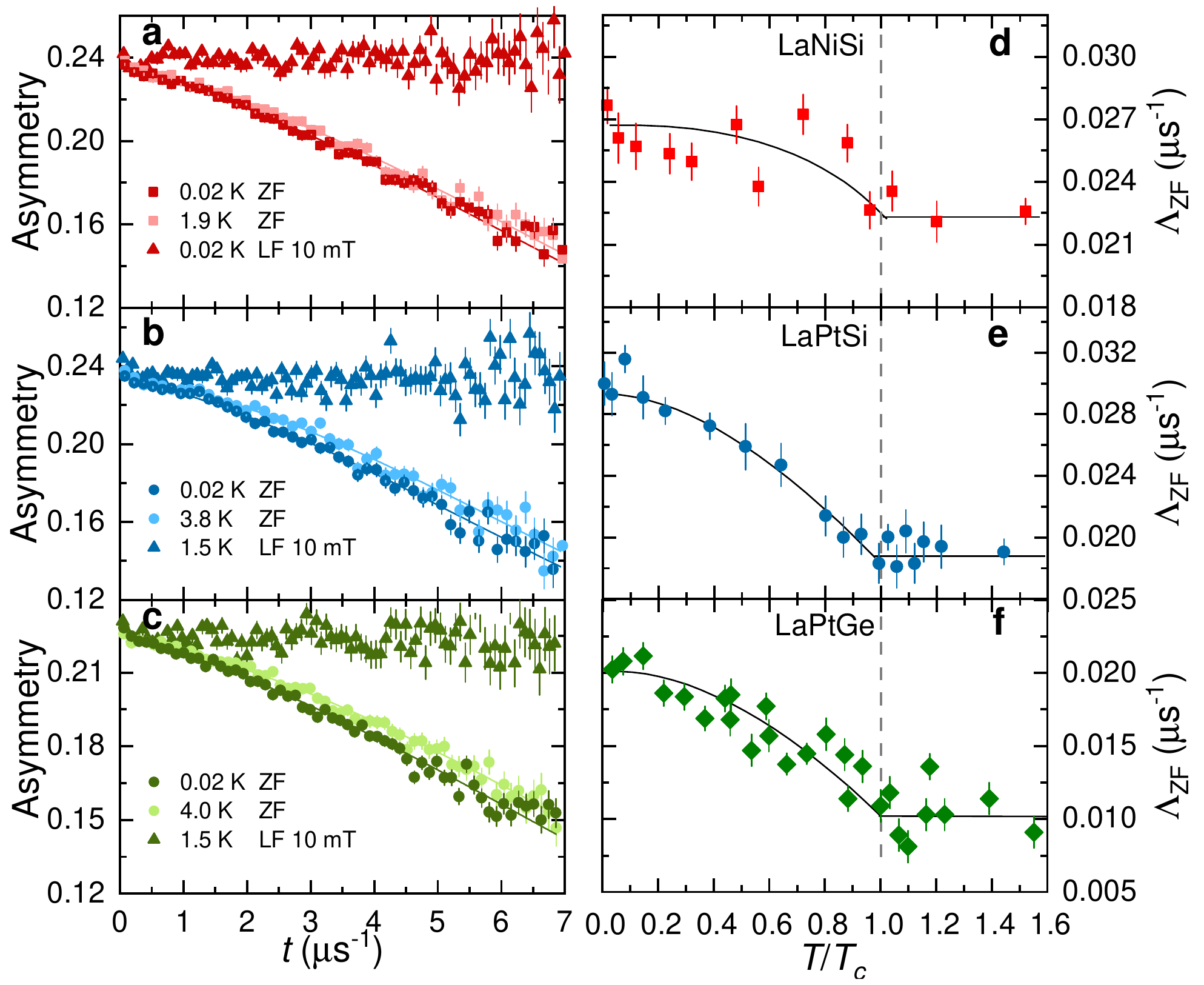}
\end{center}
	\caption{\label{fig:ZFmuSR}\textbf{Probing TRS breaking superconductivity via ZF-$\mu$SR}. \textbf{a, b, c,} Zero-field $\mu$SR spectra collected above and below $T_c$ for LaNiSi (\textbf{a}), LaPtSi (\textbf{b}), and LaPtGe (\textbf{c}). 
	In all cases, the lack of any oscillations implies a lack of long-range magnetic order.  Solid lines through the data in \textbf{a}-\textbf{c} are fits to Eq.~\eqref{eq:KT_and_electr}. The flat $\mu$SR datasets in \textbf{a}-\textbf{c} correspond to LF-$\mu$SR spectra, suggesting that even a small longitudinal field is sufficient to decouple muon spins from the local field. 
 \textbf{d, e, f,} Zero-field muon-spin relaxation rate $\Lambda_\mathrm{ZF}$ versus the reduced temperature $T/T_c$ for LaNiSi (\textbf{d}), LaPtSi (\textbf{e}), and LaPtGe (\textbf{f}). Solid lines in \textbf{d}-\textbf{f} are guides to the eyes. A consistent increase of  $\Lambda_\mathrm{ZF}$  below $T_c$ reflects the onset of spontaneous magnetic fields, indicative of a breaking of TRS in the superconducting state, while the $\sigma_\mathrm{ZF}$ is almost temperature independent (see details in Fig.~S7 in Suppl.\ Mat.). The error bars of $\Lambda_\mathrm{ZF}$ are the SDs obtained from fits to Eq.~\eqref{eq:KT_and_electr} by the \texttt{musrfit} software package~\cite{Suter2012}.}
\end{figure*}

\begin{figure*}[htb]
\begin{center}
 \includegraphics[width=0.99\textwidth]{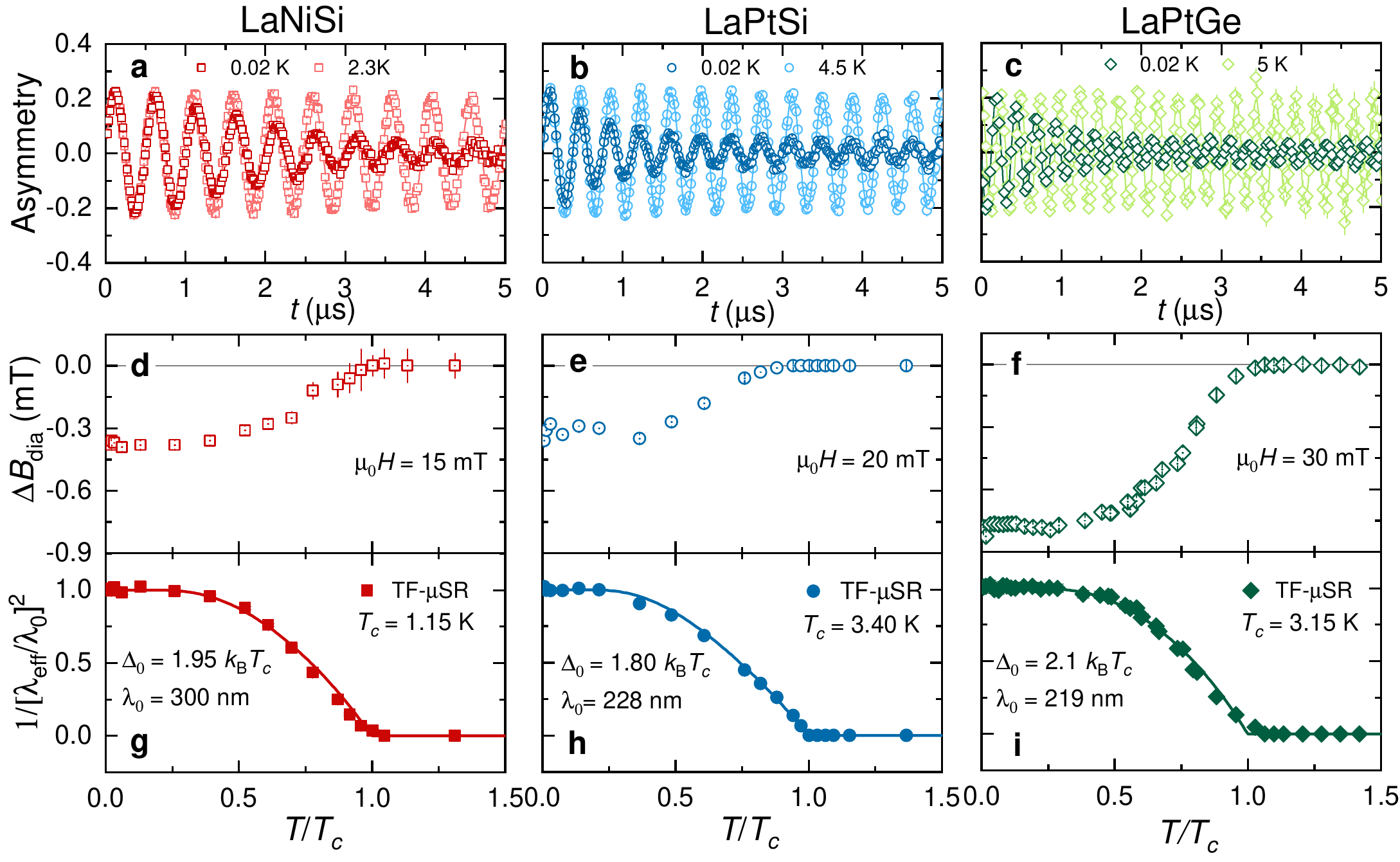}
\end{center}
\caption{\label{fig:TFmuSR}\textbf{Exploring the superconducting pairing via TF-$\mu$SR}. \textbf{a, b, c,} TF-$\mu$SR spectra, collected in the superconducting and normal states (i.e., above and below $T_c$) of LaNiSi (\textbf{a}), LaPtSi (\textbf{b}), and LaPtGe (\textbf{c}). The optimal field values for the TF-$\mu$SR measurements were identified from the lower critical field $H_\mathrm{c1}$ and the field-dependent superconducting relaxation rate (see Fig.~\ref{fig:field} and Fig.~S8 in Suppl.\ Mat.) 
and correspond to 15, 20, and 30\,mT for LaNiSi, LaPtSi, and LaPtGe, respectively. 
\textbf{d, e, f,} Diamagnetic shift $\Delta B_\mathrm{dia}$ versus the reduced temperature $T/T_c$ for LaNiSi (\textbf{d}), LaPtSi (\textbf{e}), and LaPtGe (\textbf{f}). Here, $\Delta B_\mathrm{dia} = B_\mathrm{s} - B_\mathrm{appl}$, with $B_\mathrm{s}$ the local magnetic field sensed by implanted muons in the sample
and $B_\mathrm{appl}$ the applied magnetic field. Due to the formation of the FLL, a diamagnetic field shift appears below $T_c$. 
\textbf{g, h, i,} Superfluid density [$\rho_\mathrm{sc}(T) \propto \lambda_\mathrm{eff}^{-2}(T)$] as a function of reduced temperature $T/T_c$ for LaNiSi (\textbf{g}), LaPtSi (\textbf{h}), and LaPtGe (\textbf{i}).
Solid lines represent fits to a fully-gapped $s$-wave model. The error bars of $\lambda^{-2}(T)$ are the SDs obtained from fits of the TF-$\mu$SR spectra to Eq.~\eqref{eq:TF_muSR} by the \texttt{musrfit} software package~\cite{Suter2012}.}
\end{figure*}

\noindent\textbf{Results}\\
\\
\noindent\textbf{Bulk superconductivity}\\
We synthesized three isostructural LaNiSi, LaPtSi, and LaPtGe samples 
and investigated systematically their physical properties via 
magnetic-susceptibility, specific-heat, electrical-resistivity, 
and $\mu$SR measurements. 
As shown in Fig.~\ref{fig:bulkSC}\textbf{a}, the 111-type materials crystallize 
in a noncentrosymmetric body-centered tetragonal 
structure. 
The corresponding $I4_1md$ space group (No.~109), confirmed by refinements 
of the powder x-ray diffraction (XRD) patterns (see e.g. in Fig.~\ref{fig:bulkSC}\textbf{b}), 
is nonsymmorphic and has a Bravais lattice with point group $C_{4v}$ ($4mm$). 
Upon zero-field cooling, full diamagnetic screening (i.e., bulk SC) is 
found in the magnetic susceptibility measurements in an applied field of 
1\,mT (Fig.~\ref{fig:bulkSC}\textbf{c}). Consistent with previous 
studies~\cite{Lee1994,Kneidinger2013,Evers1984}, we find $T_c = 1.28$, 
3.62, and 3.46\,K for LaNiSi, LaPtSi, and LaPtGe, respectively. 
A prominent specific-heat jump at each superconducting transition 
(see below) confirms once more the bulk SC nature of these materials.  

\vspace{5pt}
\noindent\textbf{Lower- and upper critical fields}\\
\tcr{For reliable transverse-field (TF) $\mu$SR measurements in a superconductor, 
the applied magnetic field should exceed the lower critical field $H_\mathrm{c1}$ and be much less than the upper critical field $H_\mathrm{c2}$, 
so that the additional field-distribution broadening due to the flux-line 
lattice (FFL) can be quantified from the muon-spin relaxation rate, the 
latter being directly related to the magnetic penetration depth and thus, the superfluid density.
The $H_{c1}$ values determined from field-dependent magnetization data 
are summarized in Figs.~\ref{fig:field}\textbf{a}-\textbf{c}, which provide 
lower critical fields $\mu_{0}H_{c1}(0)$ = 3.9(5), 9.6(2), and 11.8 (1)\,mT for LaNiSi, LaPtSi, and LaPtGe, respectively.
These $H_\mathrm{c1}(0)$ values 
are fully consistent with those determined from magnetic penetration 
depth (see below).
We investigated also the upper critical fields $H_{c2}$ of 111 materials, 
here shown in Figs.~\ref{fig:field}\textbf{d}-\textbf{f} 
versus the reduced temperature $T/T_c(0)$ for LaNiSi, LaPtSi, and LaPtGe, respectively.
Three different models, including Ginzburg–Landau (GL)~\cite{Zhu2008}, 
Werthamer–Helfand–Hohenberg (WHH)~\cite{Werthamer1966}, and the 
two-band model~\cite{Gurevich2011} were used to analyze the 
$H_\mathrm{c2}(T)$ data.
In LaNiSi and LaPtGe, $H_\mathrm{c2}(T)$ is well described by the WHH 
model, yielding $\mu_0 H_{c2}(0) = 0.10(1)$ and 0.44(1)\,T, respectively. 
Conversely, in LaPtSi, both WHH and GL models reproduce $H_{c2}(T)$ 
reasonably only at low fields, i.e., $\mu_0 H_\mathrm{c2} < 0.2$\,T. 
At higher magnetic fields, both models deviate significantly from the 
experimental data. Such a discrepancy most likely hints at multiple superconducting
gaps in LaPtSi, as evidenced also by the positive curvature of $H_{c2}(T)$, 
a typical feature of multigap superconductors. Indeed, here the two-band model 
shows a remarkable agreement with the experimental data and provides $\mu_0$$H_{c2}(0) = 1.17(2)$\,T. 
The presence of multiple superconducting gaps is also supported by the field-dependent electronic specific-heat coefficient (see Suppl.\ Fig.~S11).  
However, note that, due to similar gap sizes (or to small relative 
weights), the multigap features are not easily discernible 
in the superfluid density or in the zero-field electronic specific 
heat~\cite{Shang2019,Shang2020d}.
}

\vspace{5pt}
\noindent\textbf{ZF-$\mu$SR and evidence of TRS breaking}\\
Zero-field (ZF)-$\mu$SR is a very sensitive method for detecting weak magnetic fields (down to $\sim$ 0.01 mT~\cite{Amato1997}) due to the large muon gyromagnetic ratio (851.615 MHz/T) and to the availability of nearly 100\% spin-polarized muon beams. Therefore, the ZF-$\mu$SR technique has been successfully used to study different types of unconventional superconductors with broken TRS in their superconducting state~\cite{Hillier2009,Barker2015,Luke1998,Aoki2003,Shang2018a,Shang2018b,Luke1993,Shang2020a,Shang2020b}. To search for the presence of TRS breaking in the superconducting state of 111 materials, ZF-$\mu$SR measurements were performed at various temperatures, covering both their normal- and superconducting states. Representative ZF-$\mu$SR spectra are shown in Figs.~\ref{fig:ZFmuSR}\textbf{a}-\textbf{c} for LaNiSi, LaPtSi, and LaPtGe, respectively. The ZF-$\mu$SR spectra exhibit small yet clear differences between 0.02\,K and temperatures 
above $T_c$ (e.g., 1.9\,K) for LaNiSi, which become more evident in the 
LaPtSi and LaPtGe case. 

In general, in absence of external magnetic fields, the muon-spin 
relaxation is mostly determined by the interaction of muon spins with 
the randomly oriented nuclear magnetic moments. 
Thus, the ZF-$\mu$SR asymmetry can be described by means of a 
phenomenological relaxation function, consisting of a combination of 
Gaussian- and Lorentzian Kubo-Toyabe relaxations [see Eq.~\eqref{eq:KT_and_electr}]~\cite{Kubo1967,Yaouanc2011}.
While $\sigma_\mathrm{ZF}(T)$ is found to be nearly temperature 
independent (see Suppl. Fig.~S7), as shown in Figs.~\ref{fig:ZFmuSR}\textbf{d}-\textbf{f}, 
all three compounds show a clear increase of the muon-spin relaxation in the  $\Lambda_\mathrm{ZF}$ channel below $T_c$. 
Conversely, for $T > T_c$, $\Lambda_\mathrm{ZF}(T)$ is flat, thus 
excluding a possible origin 
related to magnetic impurities 
(the latter typically follow a Curie-Weiss behavior~\cite{Shang2019}).  
Furthermore, longitudinal-field (LF) $\mu$SR measurements at base 
temperature (see Figs.~\ref{fig:ZFmuSR}\textbf{a}-\textbf{c}) indicate 
that a field of only 10\,mT is sufficient to decouple the muon spins 
from the TRS breaking relaxation channel in all three compounds, indicating that the weak internal fields are static within the muon lifetime. Furthermore, the LF-$\mu$SR results 
rule out an extrinsic origin for the enhanced $\Lambda_\mathrm{ZF}(T)$. 
Considered together, the ZF- and LF-$\mu$SR results reveal that the 
increase in $\Lambda_\mathrm{ZF}(T)$ below $T_c$ is clear evidence of  
the occurrence of spontaneous magnetic fields~\cite{Hillier2009,Barker2015,Luke1998,Aoki2003,Shang2018a,Shang2018b,Luke1993,Shang2020a,Shang2020b} 
and, hence, of the \emph{breaking of TRS} in the superconducting state of LaNiSi, LaPtSi, and LaPtGe. \\

\vspace{5pt}
\noindent\textbf{TF-$\mu$SR and nodeless superconductivity}\\
To investigate the superconducting order parameter of LaNiSi, LaPtSi, 
and LaPtGe, the temperature dependence of their magnetic penetration 
depth was determined via TF-$\mu$SR 
measurements. The development of a 
flux-line lattice in the mixed state of a superconductor broadens 
the internal field distribution and leads to an enhanced muon-spin relaxation rate. 
Since the latter is determined by the magnetic penetration depth and, 
ultimately, by the superfluid density, the superconducting order parameter can be evaluated from the temperature-dependent TF-$\mu$SR measurements (see Methods). 
Following a field-cooling protocol down to  0.02\,K, the TF-$\mu$SR spectra 
were collected at various temperatures upon warming, covering both the 
superconducting and the normal states.
As shown in Figs.~\ref{fig:TFmuSR}\textbf{a}-\textbf{c}, below $T_c$, the fast decay in the TF-$\mu$SR asymmetry caused by the FLL is clearly visible. By contrast, the slow decay in the normal state, is attributed to the nuclear magnetic moments, being similar to the ZF-$\mu$SR in Figs.~\ref{fig:ZFmuSR}\textbf{a}-\textbf{c}. The TF-$\mu$SR spectra were analyzed by means of Eq.~\eqref{eq:TF_muSR}. 
Above $T_c$, the relaxation rate is small and temperature-independent, but below $T_c$ it starts to increase due to the formation of a FLL and the increased
superfluid density. At the same time, a diamagnetic field shift appears below $T_c$ (see Figs.~\ref{fig:TFmuSR}\textbf{d}-\textbf{f}).
The effective magnetic penetration depth and the superfluid density were calculated from the measured superconducting Gaussian relaxation rates 
(see Methods). 
The normalized inverse-square of the effective magnetic penetration 
depth $\lambda_\mathrm{eff}^{-2}(T)$ (proportional to the superfluid density) 
vs.\ the reduced temperature $T/T_\mathrm{c}$ for LaNiSi, LaPtSi, and LaPtGe 
is presented in Figs.~\ref{fig:TFmuSR}\textbf{g}-\textbf{i}, respectively.   
Although these three NCSCs exhibit different $T_c$ values and ASOC strengths, 
below $T_c/3$, their $\lambda_\mathrm{eff}^{-2}$ values are practically independent  of temperature.
The low-$T$ invariance of $\lambda_\mathrm{eff}^{-2}(T)$ and, 
consequently, of the superfluid density, clearly suggests a lack of 
low-energy excitations and, hence, a nodeless superconductivity 
in LaNiSi, LaPtSi, and LaPtGe, in good agreement with the low-$T$ 
electronic specific-heat data (see below) and magnetic penetration depth  
measurements via the tunnel-diode-oscillator technique~\cite{TDOnote}.	
The solid lines through the data in Figs.~\ref{fig:TFmuSR}\textbf{g}-\textbf{i} 
are fits using a fully-gapped $s$-wave model with a single superconducting gap. These 
yield gap values $\Delta_0 = 1.95(5)$, 1.80(5), and 2.10(5)\,$k_\mathrm{B}T_c$, and $\lambda_0$ = 300(3), 228(3), and 219(2)\,nm for LaNiSi, LaPtSi, and LaPtGe, respectively. 
The gap values determined from the TF-$\mu$SR measurements  
are highly consistent with those derived from specific-heat results 
(see below). \\

\begin{figure}[!h]
\centering
\includegraphics[width=0.999\columnwidth]{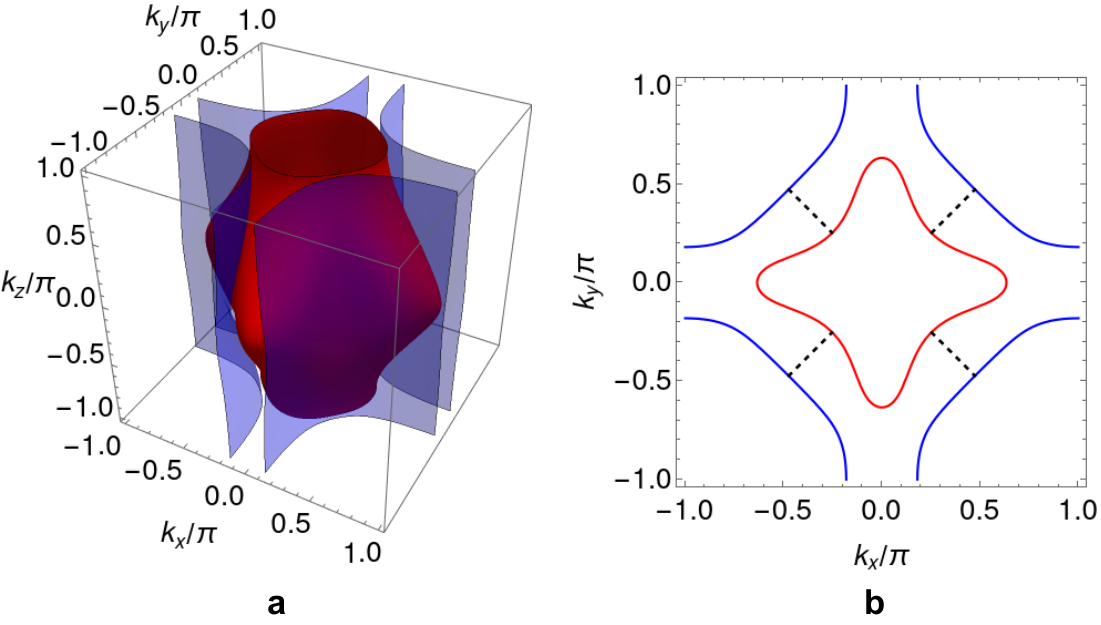}
\caption{\label{fig:FS_tbm}\textbf{Fermi surfaces for a minimal two-band model.}
\textbf{a, b,} The two important Fermi surfaces of the 111-type materials can be qualitatively reproduced by a range of parameters in a minimal two-band model. Here, we set $t_\parallel = 1$\,eV, and use the values (normalized by $t_{\parallel}$): $\mu = -1.5$, $t_{\perp} = 0.70$, $t_d = 1.25$, $t_{\delta} = 0.60$, $t_m = 0.40$, $\varepsilon^{(0)}_1 = 0.45$, and $\varepsilon^{(0)}_2 = -0.20$. \tcr{The two corresponding Fermi surfaces are shown in (\textbf{a}), 
while their projections on the $k_z = 0$ plane are shown in (\textbf{b}). 	
The dashed lines depict schematically 
the interband pairing.}}
\end{figure}

\begin{figure*}[!htb]
\begin{center}
 \includegraphics[width=1\textwidth]{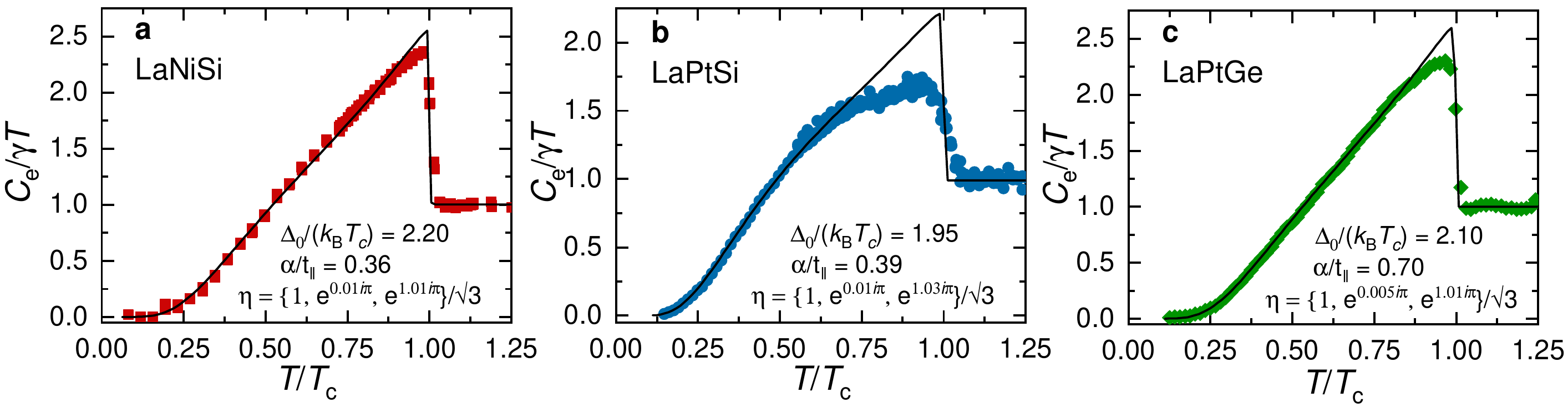}
\end{center}
\caption{\label{fig:spheat}\textbf{Electronic specific heat.} \textbf{a, b, c,} 
Normalized electronic specific heat $C_\mathrm{e}/\gamma T$ (with $\gamma$ the normal-state electronic specific-heat coefficient) as a function of reduced temperature $T/T_c$ for LaNiSi (\textbf{a}), LaPtSi (\textbf{b}), and LaPtGe (\textbf{c}). 
$C_\mathrm{e}/T$ was determined by subtracting the phonon contribution from the raw specific-heat data (see Fig.~S10 in Suppl.\ Mat.). Solid lines through the data represent theoretical calculations corresponding to the INT state with the fit  parameters listed in the figures. For LaPtSi, the reduced specific-heat jump at $T_c$ is mostly caused by the broadening of the superconducting transition.}
\end{figure*}

\vspace{5pt}
\noindent\textbf{Minimal two-band model and electronic specific heat}\\
The three 111 materials, LaNiSi, LaPtSi, and LaPtGe share similar band
structures and are inherently multiband systems, with several orbitals 
contributing to the density of states (DOS) at the Fermi level~\cite{Zhang2020}. 
The lack of an inversion center implies that an antisymmetric spin-orbit
coupling is naturally present in these materials. Here, the ASOC splits
the bands near the Fermi level, with an increasingly larger strength
from LaNiSi to LaPtSi 
to LaPtGe~\cite{Zhang2020}. In their normal state, all of them are
nonmagnetic and, thus, preserve TRS. These materials have been predicted
to exhibit four Weyl nodal rings around the $X$ point, at $\sim 0.5$\,eV
below the Fermi level, a topological feature protected by nonsymmorphic
glide mirror symmetry and TRS~\cite{Zhang2020}. The size of the nodal
rings increases with increasing ASOC strength and, 
due to their presence, 
the 111-type Weyl nodal-line semimetals are expected to show interesting
magneto-transport properties.

The antisymmetric spin-orbit coupling, however, does not change the
topology of Fermi surfaces qualitatively.
In 111 materials, in the absence of ASOC, three spin-degenerate bands
cross the Fermi level giving rise to three Fermi surfaces with similar 
shapes.
However, only two of them contribute significantly ($\sim 96\%$) to the
DOS at the Fermi level~\cite{Zhang2020}. 
The low-energy properties of 111 materials are thus dominated by these two
Fermi surfaces.
To capture qualitatively their topology, we build a minimal two-band
tight-binding model by suitably choosing the chemical potential. 
The dispersions of the two bands are:
\beq
\epsilon_j(\bk) = \varepsilon^{(0)}_j + g_1(\bk) + (-1)^j g_2(\bk),
\eeq 
where $j = 1$, 2; $\varepsilon^{(0)}_j$ are the onsite energies,
$g_1(\bk) = [\epsilon'(\bk) + \epsilon''(\bk)]/2$, $g_2(\bk) = [\{\epsilon'(\bk) - \epsilon''(\bk)\}^2/4 + t^2_m]^{1/2}$, with $\epsilon'(\bk) = [\epsilon(k_x)+\epsilon(k_y)]/2-[\{\epsilon(k_x) - \epsilon(k_y)\}^2/4 + t^2_\delta]^{1/2}$, $\epsilon(x)=-2t_{\parallel} \cos(x)$, $\epsilon''(\bk) = -2t_{\perp} \cos(k_z) - 2t_d [\cos(k_x)+\cos(k_y)]$. Here, $t_{\parallel}$, $t_m$, $t_d$, $t_{\delta}$, and $t_{\perp}$ are the hopping parameters. The corresponding Fermi surfaces for a realistic 
choice of the parameters --- to be used in the subsequent discussion ---
are shown in Fig.~\ref{fig:FS_tbm}\textbf{a}.

To account for the ASOC effects in the minimal two-band model, we
note that the form of ASOC for the corresponding $C_{4v}$ point
group is $V_\mathrm{ASOC} = \alpha_{xy} (k_y \sigma_x - k_x \sigma_y) + \alpha_z k_x k_y k_z (k^2_x - k^2_y)\sigma_z$, where $\pmb{\sigma} = (\sigma_x, \sigma_y, \sigma_z)$ is the vector of Pauli matrices in the spin space, while $\alpha_{xy}$ and $\alpha_z$ are the
strengths of the two types of ASOC terms allowed by symmetry~\cite{smidman2017}. Note that,
the second term is of fifth order in $k$ and leads to spin splitting.
On the other hand, the Rashba term, with strength $\alpha_{xy}$, is expected
to be dominant because of the quasi-2D nature of the two Fermi surfaces
(see Fig.~\ref{fig:FS_tbm}\textbf{a}). Hence, only the Rashba ASOC term is
phenomenologically relevant
in the minimal model. 
In general, this term would have both interband- and intraband contributions.
However, to keep the topology of the Fermi surfaces with ASOC similar
to that without ASOC and to correctly describe the experimental observations in
the 111 materials, we need to work in the limit where the \emph{interband}
contribution is
large compared to the intraband one 
(see Suppl.\ Mat.\ for details). This emphasizes the interband nature
of the pairing under consideration. As a result, we only consider an
interband Rashba ASOC of strength $\alpha$, i.e., $V^\mathrm{inter}_\mathrm{R} = \alpha (k_y \sigma_x - k_x \sigma_y)$. The normal-state Hamiltonian then takes the form
$\hat{\cal{H}}_{\rm N} = \sum_{\bk} \hat{c}^{\dagger}_{\bk} \cdot H_{\rm N}(\bk) \cdot \hat{c}_{\bk}$, where $H_{\rm N}(\bk) = \sigma_0 \otimes \begin{bmatrix} \xi_1(\bk) & 0 \\ 0 & \xi_2(\bk) \end{bmatrix} + \alpha (k_y \sigma_x - k_x \sigma_y) \otimes \tau_x$. Here, $\xi_j(\bk) = \epsilon_j(\bk) - \mu$,
with $\mu$ being the chemical potential, $\pmb{\tau} = (\tau_x, \tau_y, \tau_z)$ is the vector of Pauli matrices in the band space, $\sigma_0$ and $\tau_0$ are the identity matrices in the spin and band spaces,  respectively. Further, $\hat{c}_{\bk} = \begin{bmatrix} \tilde{c}_{\uparrow, \bk} \\ \tilde{c}_{\downarrow, \bk} \end{bmatrix}$, with $\tilde{c}_{s,\bk} = \begin{bmatrix} c_{1, s, \bk} \\ c_{2, s, \bk} \end{bmatrix}$, where $c_{m, s, \bk}$ is a fermion annihilation operator in the band
$m = 1,2$ with spin $s = \uparrow, \downarrow$. 

Due to the inherent multiband nature and to the presence of nonsymmorphic
symmetries, the usual classification of the possible superconducting order
parameters (based on point-group symmetries) in the effective single-band picture is insufficient for the 111 materials (see Suppl.\ Mat.\ for details).
Indeed, nonsymmorphic symmetries can lead to additional symmetry-imposed 
nodes along the high symmetry directions on the zone faces~\cite{sumita2018,sumita2019}. 
\tcr{Even a loop supercurrent state~\cite{Ghosh2021}, which has a uniform
onsite singlet pairing and proposed to be realized in some of the fully-gapped TRS breaking superconductors,  is not allowed in the case of 111 materials, because there are only two symmetrically distinct sites within a unit cell}. However, we note that the two Fermi surfaces under consideration have large sections in the Brillouin zone which are almost parallel and close to each other (see Fig.~\ref{fig:FS_tbm}\textbf{b}). Hence, to consistently explain the
phenomenon of TRS breaking at $T_c$ with the presence of a full SC gap,
we expect that an internally antisymmetric nonunitary triplet (INT)
state~\cite{Weng2016,Ghosh2020b}, which features a uniform pairing
between same spins in the two different bands, to become the dominant
instability. In this state, the pairing potential matrix is $\hat{\Delta} = \hat{\Delta}_S \otimes \hat{\Delta}_B$, where $\hat{\Delta}_S$ and $\hat{\Delta}_B$ are the pairing
potential matrices in the spin- and band spaces, respectively. $\hat{\Delta}_B = i \tau_y$ gives the required fermionic antisymmetry. $\hat{\Delta}_S = (\bd.\pmb{}\sigma)i\sigma_y$,
where $\bd = \Delta_0 \pmb{\eta}$, with $|\pmb{\eta}|^2 = 1$, is the $\bd$-vector characterizing the triplet pairing state, which is nonunitary because $\bq = i (\pmb{\eta} \times \pmb{\eta}^*) \neq 0$. $\Delta_0$ is an overall pairing amplitude.

We compute the Bogoliubov quasiparticle energies $E_n(\bk)$, $n=1,\ldots 4$
for the effective model in the INT ground state using the Bogoliubov-de-Gennes (BdG)
formalism (see Suppl.\ Mat.\ for details). The thermodynamic properties
are computed by assuming that the temperature dependence comes only from
the pairing amplitude in the form 
$\Delta(T) = \Delta_0 \mathrm{tanh} \{ 1.82[1.018(T_\mathrm{c}/T-1)]^{0.51} \}$
and ignoring any weak temperature dependence of
the $\bq$-vector~\cite{Ghosh2022}. To reproduce the experimental
specific-heat results for the three materials, three fitting parameters,
namely, $\Delta_0/(k_\mathrm{B} T_c)$, the direction of $\bd$-vector,
and $\alpha$ had to be tuned to get the best fits in the weak\--coupling
limit (see Fig.~\ref{fig:spheat}).
Note that, for all the three materials we can reproduce 
the specific-heat data
rather well \tcr{(especially at low temperatures)} and the fitting process naturally preserves the trend of increasing ASOC strength in the 111 family. 
More importantly, the derived superconducting energy gaps are highly consistent with the values determined from TF\--$\mu$SR data. \tcr{The nonzero real vector $\bq$, 
found from the fits, 
points in different directions for the three materials and encodes the effective TRS\--breaking field arising from spin-polarization caused by Cooper-pair migration due to the nonunitary nature of 
pairing~\cite{Ghosh2020b}.}\\

\noindent\textbf{Discussion}\\

According to ZF-$\mu$SR results in 111 materials, the spontaneous
magnetic fields or the magnetization in the superconducting state of
LaPtSi or LaPtGe are much larger than in LaNiSi, \tcr{here reflected 
in significantly larger variations of $\Lambda_\mathrm{ZF}$ between 
zero-temperature and $T_c$ in the former two cases as compared to LaNiSi.
Therefore, the TRS breaking effect is more prominent 
in the superconducting state of LaPtSi and LaPtGe than of 
LaNiSi (see Fig.~\ref{fig:ZFmuSR}).} 
Previous ZF-$\mu$SR studies indicate that, although LaNiSi and LaPtSi
exhibit an enhanced muon-spin relaxation rate at low temperatures,
their $\Lambda_\mathrm{ZF}(T)$ resembles a Curie-Weiss behavior [i.e., $\lambda_\mathrm{ZF}(T)$ $\propto$ $T^{-1}$]. \tcr{This, and the lack of a distinct anomaly 
in $\Lambda_\mathrm{ZF}(T)$ near $T_c$~\cite{Sajilesh2020}, are 
inconsistent with the TRS breaking effect. In general, an enhanced 
muon-spin relaxation with Curie-Weiss feature might be related to either intrinsic- or to 
extrinsic spin fluctuations. 
As for the intrinsic case, a typical example is that of the ThFeAsN 
iron-based superconductor. It exhibits strong magnetic fluctuations 
at low temperatures (confirmed also by nuclear magnetic resonance measurements), 
which are reflected in a steadily increasing $\Lambda_\mathrm{ZF}(T)$ 
as the temperature is lowered~\cite{Shiroka2017}.
As for the extrinsic case, a typical example is that of the ReBe$_{22}$ 
multigap superconductor. Here, $\Lambda_\mathrm{ZF}(T)$ increases remarkably 
with decreasing temperature due to the tiny amounts of magnetic 
impurities, whose contribution is enhanced near zero temperature~\cite{Shang2019}. 
Conversely, in case of a truly broken TRS -- for instance, in 111 materials 
we report here --} 
$\Lambda_\mathrm{ZF}$ is almost independent of
temperature for $T > T_c$, 
strongly suggesting that the 
enhanced $\Lambda_\mathrm{ZF}$ is induced by the \emph{spontaneous fields} 
occurring in the superconducting state. 

For LaPtGe, previous ZF-$\mu$SR data show similar
features to our results (see Fig.~\ref{fig:ZFmuSR}f), i.e., a small yet
clear difference in the ZF-$\mu$SR spectra between 0.3 and 4.5\,K~\cite{Sajilesh2018},
the latter dataset referring to the normal state. \tcr{However, the authors claimed that, the temperature-dependent $\sigma_\mathrm{ZF}$ and $\Lambda_\mathrm{ZF}$ 
exhibit no visible differences and, thus, a preserved TRS was concluded~\cite{Sajilesh2018}.} 
By contrast, our systematic ZF-$\mu$SR measurements
suggest the presence of spontaneous magnetic fields, hence, the broken
TRS in the superconducting state of LaPtGe.
\tcr{Such discrepancies in ZF-$\mu$SR results might be related to the 
different sample \emph{quality}, \emph{purity}, or \emph{disorder}. For example, the previous 
study reports a residual resistivity $\rho_0 \sim 200$\,$\mu$$\Omega$cm~\cite{Sajilesh2018}, 
three times larger than that of current  LaPtGe sample, $\rho_0$ $\sim$ 66\,$\mu$$\Omega$cm [see Suppl. Fig.~S2]. 
Moreover, the residual resistivity ratio of the current LaPtGe sample is twice larger than that of the previous sample.} 
Nevertheless, to independently confirm the TRS breaking in the 
superconducting state of 111 materials, the use of other techniques, 
\tcr{as e.g., Josephson tunneling, SQUID, or optical Kerr effect, is highly desirable. In particular, the optical Kerr effect, another very sensitive probe of spontaneous fields in unconventional superconductors, is renown for confirming TRS breaking in Sr$_2$RuO$_4$ and UPt$_3$~\cite{Xia2006,Schemm2014}.
In addition, to exclude disorder effects, search for possible non-$s$-wave behavior, 
and confirm the TRS breaking in 111 materials, in the future, 
measurements on high-quality single crystals will clearly be helpful.}
	
%
\tcr{According to the Uemura plot~\cite{Uemura1991}, clearly, the 111 materials studied here lie in the TRS-breaking band, where different families of superconductors are found to break the TRS in the superconducting state (see details in Supplementary Figure 16 and Note 10).}
Apart from the La-based 111 materials studied here, also the isostructural 
Th$T$Si compounds (with $T$ = Co, Ni, Ir, and Pt) are superconductors 
(with critical temperatures between 2 and 6.5\,K)~\cite{Domieracki2018,Zhong1985}. 
Similar to the La-based cases, 
the Th-based materials, too, exhibit a large ASOC upon replacing the 3$d$ 
Ni and Co with 5$d$ Pt and Ir~\cite{Ptok2019}. Recently, superconductivity 
with $T_c = 5.07$\,K was reported in ThIrP, which also adopts a LaPtSi-type 
structure~\cite{Xiao2021}. Therefore, it would be interesting to search 
for possible TRS breaking and, hence, unconventional superconductivity 
in these Th-based 111 materials. In addition, La-based 111 materials, 
in particular LaNi$_{1-x}$Pt$_x$Si, represent ideal candidate systems 
for investigating the effect of ASOC on spontaneous magnetization and 
unconventional superconductivity.

Generally, in noncentrosymmetric superconductors, the ASOC can induce 
a mixing of singlet- and triplet states. However, in the 111 materials 
under consideration, it plays a crucial role in stabilizing even a 
purely triplet state. Moreover, the necessity of a dominant interband contribution to the ASOC
	 in achieving a fully gapped spectrum in the INT state, further justifies the interband pairing 
	 in the superconducting state.
We also note that the triplet 
$\bd$-vectors, obtained from analyses of the specific-heat data, 
correspond to a partially spin-polarized ($|\bq| < 1 \neq 0$) 
superconducting state. In this case, the spontaneous magnetization 
results from a migration of Cooper pairs from the majority to a minority 
spin species~\cite{Ghosh2020b}.  

The normal state of 111 materials has a non-trivial topology due to 
the Weyl nodal lines protected by nonsymmorphic glide symmetry and TRS. 
Apart from the usual photoemission studies~\cite{Armitage2018,Lv2021}, 
the corresponding drumhead surface states can also be investigated 
by inspecting 
the correlation effects on the surfaces~\cite{Liu2017}. 
Since TRS is spontaneously broken at $T_c$, it is of interest to 
investigate the fate of the bulk nodal lines. 
Our results demonstrate that 111 materials represent a rare case of 
Weyl nodal-line semimetals which break time-reversal symmetry in the 
superconducting state. As such, they epitomize 
the ideal system for investigating the rich interplay between the 
exotic properties of topological nodal-line fermions and unconventional 
superconductivity.

\vspace{10pt}
\noindent\textbf{Methods}\\
\\
\textbf{Sample preparation.} Polycrystalline LaNiSi, LaPtSi, and 
LaPtGe samples were prepared by arc melting La (99.9\%, Alfa Aesar), 
Ni (99.98\%, Alfa Aesar), Pt (99.9\%, ChemPUR), Si (99.9999\%, 
Alfa Aesar) and Ge (99.999\%, Alfa Aesar) in high-purity argon 
atmosphere. To improve homogeneity, the ingots were flipped and re-melted 
more than five times. The as-cast ingots were then annealed at 800$^\circ$C 
for two weeks. 
The crystal structure and purity of the samples were checked using powder 
x-ray diffraction at room temperature using a Bruker D8 diffractometer with Cu K$_\alpha$ radiation.  
All three compounds crystallize in a tetragonal noncentrosymmetric structure 
with a space group $I4_1md$ (No.\,109). The estimated lattice parameters 
are listed in the Suppl.\ Table S1.\\

\noindent\textbf{Sample characterization.} 
The magnetization, heat-capacity, and electrical-resistivity measurements were performed 
on a Quantum Design magnetic property measurement system (MPMS) and 
a physical property measurement system (PPMS). The lower critical field 
$H_\mathrm{c1}$ was determined by field-dependent magnetization measurements 
at various temperatures up to $T_\mathrm{c}$, while the upper critical field 
$H_\mathrm{c2}$ was determined by measuring the temperature-dependent 
heat capacity under various magnetic fields, 
and by field-dependent magnetization at various temperatures.\\

\noindent\textbf{$\mu$SR experiments.} 
The $\mu$SR experiments were conducted at the general-purpose surface-muon
(GPS) and at the low-temperature facility (LTF) instruments of the Swiss muon 
source (S$\mu$S) at Paul Scherrer Institut (PSI) in Villigen, Switzerland. 
Once implanted in a material, at a typical depth of $\sim 0.3$\,mm, the 
spin-polarized positive muons ($\mu^+$) act as microscopic probes of the 
local magnetic environment via the decay positrons, emitted preferentially 
along the muon-spin direction.
The spatial anisotropy of the emitted positrons 
(i.e., the asymmetry signal) reveals the distribution of the local magnetic 
fields at the muon stopping sites~\cite{Blundell1999,Yaouanc2011}.
For TF-$\mu$SR measurements, the applied magnetic field is perpendicular 
to the muon-spin direction, while for LF-$\mu$SR measurements, the magnetic 
field is parallel to the muon-spin direction. 
In both the TF- and LF-$\mu$SR cases, the samples were cooled in an applied 
magnetic field down to the base temperature (1.5 K for GPS and 0.02 K for LTF). 
Field cooling reduces flux pinning and ensures an almost ideal 
flux-line lattice.
The $\mu$SR spectra were then collected upon heating. 
For the ZF-$\mu$SR measurements, to exclude the possibility of stray 
magnetic fields, the magnets were quenched before 
the measurements, and an active field-nulling facility was used to 
compensate for stray fields down to 1\,$\mu$T. 
\\


\noindent\textbf{Analysis of the $\mu$SR spectra.}	
All the $\mu$SR data were analyzed by means of the \texttt{musrfit} 
software package~\cite{Suter2012}. In absence of applied external 
fields, in the nonmagnetic LaNiSi, LaPtSi, and LaPtGe, the relaxation 
is mainly determined by the randomly oriented nuclear magnetic moments. 
Therefore, their ZF-$\mu$SR spectra can be modeled by means of a 
phenomenological relaxation function, consisting of a combination of 
Gaussian- and Lorentzian Kubo-Toyabe relaxations~\cite{Kubo1967,Yaouanc2011}:
\begin{equation}
	\label{eq:KT_and_electr}
	A_\mathrm{ZF} = A_\mathrm{s}\left[\frac{1}{3} + \frac{2}{3}(1 -
	\sigma_\mathrm{ZF}^{2}t^{2} - \Lambda_\mathrm{ZF} t)\,
	\mathrm{e}^{\left(-\frac{\sigma_\mathrm{ZF}^{2}t^{2}}{2} - \Lambda_\mathrm{ZF} t\right)} \right] + A_\mathrm{bg}.
\end{equation}
Here $A_\mathrm{s}$ and $A_\mathrm{bg}$ represent the initial muon-spin 
asymmetries for muons implanted in the sample and the sample holder, respectively. 
The $\sigma_\mathrm{ZF}$ and $\Lambda_\mathrm{ZF}$ represent the zero-field Gaussian 
and Lorentzian relaxation rates, respectively. Since $\sigma_\mathrm{ZF}$ shows 
an almost temperature-independent behavior, the $\Lambda_\mathrm{ZF}$ values 
in Figs.~\ref{fig:ZFmuSR}\textbf{a}-\textbf{c} could be derived by fixing 
$\sigma_\mathrm{ZF}$ to its average value, i.e., $\sigma_\mathrm{ZF}^\mathrm{av} = 0.094$, 
0.103, and 0.104\,$\mu$s$^{-1}$ for LaNiSi, LaPtSi, and LaPtGe, respectively 
(see details in Suppl.\ Fig.~S7).

In the TF-$\mu$SR case, the time evolution of the asymmetry was modeled by: 
\begin{equation}
	\label{eq:TF_muSR}
	A_\mathrm{TF}(t) =  A_\mathrm{s} \cos(\gamma_{\mu} B_\mathrm{s} t + \phi) e^{- \sigma^2 t^2/2} +
	A_\mathrm{bg} \cos(\gamma_{\mu} B_\mathrm{bg} t + \phi).
\end{equation}
Here $A_\mathrm{s}$ and $A_\mathrm{bg}$ are the same as in ZF-$\mu$SR. 
$B_\mathrm{s}$ and $B_\mathrm{bg}$ are the local fields sensed by 
implanted muons in the sample and the sample holder (i.e., silver plate), 
$\gamma_{\mu}$ is the muon gyromagnetic ratio, $\phi$ is the shared initial phase, and $\sigma$ 
is a Gaussian relaxation rate reflecting the field distribution inside the sample. 
In the superconducting state, $\sigma$ includes contributions from both 
the flux-line lattice (FLL) ($\sigma_\mathrm{sc}$) and a smaller, 
temperature-independent relaxation, due to the nuclear moments 
($\sigma_\mathrm{n}$, similar to $\sigma_\mathrm{ZF}$). The former can 
be extracted by subtracting the $\sigma_\mathrm{n}$ in quadrature, i.e., 
$\sigma_\mathrm{sc}$ = $\sqrt{\sigma^{2} - \sigma^{2}_\mathrm{n}}$. 
Since $\sigma_\mathrm{sc}$ is directly related to the effective magnetic 
penetration depth and, thus, to the superfluid density 
($\sigma_\mathrm{sc} \propto 1/\lambda_\mathrm{eff}^2 \sim \rho_\mathrm{sc}$), 
the superconducting gap and its symmetry can be investigated by measuring 
the temperature-dependent $\sigma_\mathrm{sc}$.

The effective penetration depth $\lambda_\mathrm{eff}$ had to be calculated 
from $\sigma_\mathrm{sc}$ by considering the overlap of vortex cores~\cite{Brandt2003}:
\begin{equation}
	\label{eq:sigma_to_lambda}
	\sigma_\mathrm{sc} (h) = 0.172 \frac{\gamma_{\mu} \Phi_0}{2\pi}(1-h)[1+1.21(1-\sqrt{h})^3]\lambda^{-2}_\mathrm{eff}.
\end{equation}
%
%
Here, $h = H_\mathrm{appl}/H_\mathrm{c2}$, is the reduced magnetic field, where $H_\mathrm{appl}$ represents the applied external magnetic field and $H_\mathrm{c2}$ the upper critical fields. 
Values of the latter are reported in Suppl.\ Figs.~S4-S6. 
For the TF-$\mu$SR measurements we used $\mu_0H_\mathrm{appl} = 15$, 
20, and 30\,mT for LaNiSi, LaPtSi, and LaPtGe, respectively. 
Further details about the data analysis can be found in the Supplementary Material.\\

\noindent\textbf{Superconducting gap symmetry.} Since the 111 materials exhibit almost temperature independent superfluid density below 1/3$T_c$,
to extract the superconducting gap, the temperature-dependent superfluid density $\rho_\mathrm{sc}(T)$
of LaNiSi, LaPtSi, and LaPtGe was analyzed by using a fully-gapped $s$-wave model, generally described by: 
\begin{equation}
	\label{eq:rhos}
	\rho_\mathrm{sc}(T) = \frac{\lambda_0^2}{\lambda_\mathrm{eff}^2(T)} =  1 + 2\int^{\infty}_{\Delta(T)} \frac{E}{\sqrt{E^2-\Delta^2(T)}} \frac{\partial f}{\partial E} \mathrm{d}E, 
\end{equation}
with $f = (1+e^{E/k_\mathrm{B}T})^{-1}$ the Fermi function~\cite{Tinkham1996,Prozorov2006} 
and $\lambda_0$ the effective magnetic penetration depth at zero temperature. 
The temperature evolution of the superconducting energy gap follows 
$\Delta(T) = \Delta_0 \mathrm{tanh} \{ 1.82[1.018(T_\mathrm{c}/T-1)]^{0.51} \}$, 
where $\Delta_0$ is the gap value at zero temperature~\cite{Carrington2003}.\\

\noindent\textbf{Theoretical analysis.} 
We use the Bogoliubov\--de\--Gen\-nes 
formalism
to compute the quasiparticle energy 
bands for the minimal two-band model in the INT state~\cite{Sigrist1991}. Here, we 
assume the temperature dependence comes only from the pairing amplitude. 
The Bogoliubov quasiparticle energy bands $E_n(\bk)$; $n = 1 \ldots 4$ are used to compute the specific heat by the formula:
\beq
C =\sum_{n,\bk} \frac{k_\mathrm{B} \beta^2}{2} \left[E_n(\bk) + 
\beta \frac{\partial E_n(\bk)}{\partial \beta}\right] E_n(\bk) \,\,{\rm sech}^2\left[\frac{\beta E_n(\bk)}{2}\right]
\eeq 
where $\beta = \frac{1}{k_\mathrm{B}T}$ and $k_\mathrm{B}$ is the Boltzmann constant. 
Further details about the theoretical analysis can be found in the Supplementary Material.

\vspace{10pt}
\noindent\textbf{Data availability.} 
All the data needed to evaluate the reported conclusions 
are presented in the paper and/or in the Supplementary Material. 
Additional data related to this paper may be requested from the 
authors. The $\mu$SR data
were generated at the S$\mu$S  (Paul Scherrer Institut, Switzerland).
Derived data supporting the results of this study are available from the corresponding authors or beamline scientists. The
\texttt{musrfit} software package is available online free of charge at http://lmu.web.psi.ch/musrfit/technical/index.html.

\vspace{10pt}
\noindent\textbf{Acknowledgments}\\
We acknowledge the allocation of beam time at S$\mu$S (GPS and LTF spectrometers). We thank Peiran Zhang and Chao Cao for sharing the band structure data for LaNiSi, and other related discussions.
T.S.\ acknowledges support by the Natural Science Foundation of Shanghai 
(Grant Nos.\ 21ZR1420500 and 21JC1402300). S.K.G., J.F.A.\ and J.Q.\ acknowledge 
support by EPSRC through the project ``Unconventional Superconductors: New paradigms 
for new materials'' (Grant Refs.\ EP/P00749X/1 and EP/P007392/1). S.K.G.\ also 
acknowledges the Leverhulme Trust for support through the Leverhulme early-career 
fellowship. This work was also supported by the Swiss National Science Foundation 
(Grant Nos.\ 200021\_188706, 200021\_169455, 206021\_139082). M.S.\ and H.Q.Y.\ acknowledge 
the National Key R\&D Program of China (Grant Nos.\ 2017YFA0303100 and 2016YFA0300202), 
the National Natural Science Foundation of China (Grant Nos.\ 11874320, 12034017, 
and 11974306), and the Key R\&D Program of Zhejiang Province, China 
(Grant No.\ 2021C01002).
\\

\vspace{10pt}
\noindent\textbf{Additional information}\\
\textbf{Supplementary Information} is available for this paper at xxx.\\
\textbf{Correspondence} and requests for materials should be addressed to T.S.

\vspace{10pt}
\noindent\textbf{Author contributions} \\
T.S., S.K.G., M.S., H.Q.Y., J.Q., and T.Sh. conceived and led the project. T.S., D.J.G., and E.P. synthesized the samples. T.S., T.Sh., and C.B. performed the $\mu$SR measurements. 
S.K.G. performed the theoretical analysis with advice from J.Q. and J.F.A. T.S., A.W., W.X., Y.C., M.O.A., M.N., and M.M. measured the electrical resistivity, heat capacity, and magnetization. T.S. analyzed all the experimental data. T.S., S.K.G., and M.S. wrote the paper with input from all the co-authors.

\vspace{10pt}
\noindent\textbf{Competing financial interests}\\
The Authors declare no Competing Financial or Non-Financial Interests

\vspace{10pt}
\noindent\textbf{References}
\bibliographystyle{naturemag}
\bibliography{refs_La111.bib}

\begin{thebibliography}{10}
\expandafter\ifx\csname url\endcsname\relax
  \def\url#1{\texttt{#1}}\fi
\expandafter\ifx\csname urlprefix\endcsname\relax\def\urlprefix{URL }\fi
\providecommand{\bibinfo}[2]{#2}
\providecommand{\eprint}[2][]{\url{#2}}

\bibitem{Armitage2018}
\bibinfo{author}{Armitage, N.~P.}, \bibinfo{author}{Mele, E.~J.} \&
  \bibinfo{author}{Vishwanath, A.}
\newblock \bibinfo{title}{Weyl and {Dirac} semimetals in three-dimensional
  solids}.
\newblock \emph{\bibinfo{journal}{Rev. Mod. Phys.}}
  \textbf{\bibinfo{volume}{90}}, \bibinfo{pages}{015001}
  (\bibinfo{year}{2018}).

\bibitem{Lv2021}
\bibinfo{author}{Lv, B.~Q.}, \bibinfo{author}{Qian, T.} \&
  \bibinfo{author}{Ding, H.}
\newblock \bibinfo{title}{Experimental perspective on three-dimensional
  topological semimetals}.
\newblock \emph{\bibinfo{journal}{Rev. Mod. Phys.}}
  \textbf{\bibinfo{volume}{93}}, \bibinfo{pages}{025002}
  (\bibinfo{year}{2021}).

\bibitem{Zhang2020}
\bibinfo{author}{Zhang, P.}, \bibinfo{author}{Yuan, H.} \&
  \bibinfo{author}{Cao, C.}
\newblock \bibinfo{title}{Electron-phonon coupling and nontrivial band topology
  in noncentrosymmetric superconductors {LaNiSi}, {LaPtSi}, and {LaPtGe}}.
\newblock \emph{\bibinfo{journal}{Phys. Rev. B}}
  \textbf{\bibinfo{volume}{101}}, \bibinfo{pages}{245145}
  (\bibinfo{year}{2020}).

\bibitem{Lee1994}
\bibinfo{author}{Lee, W.~H.}, \bibinfo{author}{Yang, F.~A.},
  \bibinfo{author}{Shih, C.~R.} \& \bibinfo{author}{Yang, H.~D.}
\newblock \bibinfo{title}{Crystal structure and superconductivity in the
  {Ni}-based ternary compound {LaNiSi}}.
\newblock \emph{\bibinfo{journal}{Phys. Rev. B}} \textbf{\bibinfo{volume}{50}},
  \bibinfo{pages}{6523--6525} (\bibinfo{year}{1994}).

\bibitem{Kneidinger2013}
\bibinfo{author}{Kneidinger, F.} \emph{et~al.}
\newblock \bibinfo{title}{Synthesis, characterization, electronic structure,
  and phonon properties of the noncentrosymmetric superconductor {LaPtSi}}.
\newblock \emph{\bibinfo{journal}{Phys. Rev. B}} \textbf{\bibinfo{volume}{88}},
  \bibinfo{pages}{104508} (\bibinfo{year}{2013}).

\bibitem{Evers1984}
\bibinfo{author}{Evers, J.}, \bibinfo{author}{Oehlinger, G.},
  \bibinfo{author}{Weiss, A.} \& \bibinfo{author}{Probst, C.}
\newblock \bibinfo{title}{Supraconductivity of {LaPtSi} and {LaPtGe}}.
\newblock \emph{\bibinfo{journal}{Solid State Commun.}}
  \textbf{\bibinfo{volume}{50}}, \bibinfo{pages}{61--62}
  (\bibinfo{year}{1984}).

\bibitem{Annett1990}
\bibinfo{author}{Annett, J.~F.}
\newblock \bibinfo{title}{Symmetry of the order parameter for high-temperature
  superconductivity}.
\newblock \emph{\bibinfo{journal}{Adv. Phys.}} \textbf{\bibinfo{volume}{39}},
  \bibinfo{pages}{83--126} (\bibinfo{year}{1990}).

\bibitem{Sigrist1991}
\bibinfo{author}{Sigrist, M.} \& \bibinfo{author}{Ueda, K.}
\newblock \bibinfo{title}{Phenomenological theory of unconventional
  superconductivity}.
\newblock \emph{\bibinfo{journal}{Rev. Mod. Phys.}}
  \textbf{\bibinfo{volume}{63}}, \bibinfo{pages}{239--311}
  (\bibinfo{year}{1991}).

\bibitem{Ghosh2020a}
\bibinfo{author}{Ghosh, S.~K.} \emph{et~al.}
\newblock \bibinfo{title}{Recent progress on superconductors with time-reversal
  symmetry breaking}.
\newblock \emph{\bibinfo{journal}{J. Phys: Cond. Matt.}}
  \textbf{\bibinfo{volume}{33}}, \bibinfo{pages}{033001}
  (\bibinfo{year}{2020}).

\bibitem{Hillier2009}
\bibinfo{author}{Hillier, A.~D.}, \bibinfo{author}{Quintanilla, J.} \&
  \bibinfo{author}{Cywinski, R.}
\newblock \bibinfo{title}{Evidence for time-reversal symmetry breaking in the
  noncentrosymmetric superconductor {LaNiC$_{2}$}}.
\newblock \emph{\bibinfo{journal}{Phys. Rev. Lett.}}
  \textbf{\bibinfo{volume}{102}}, \bibinfo{pages}{117007}
  (\bibinfo{year}{2009}).

\bibitem{Barker2015}
\bibinfo{author}{Barker, J. A.~T.} \emph{et~al.}
\newblock \bibinfo{title}{Unconventional superconductivity in
  {La}$_{7}${Ir}$_{3}$ revealed by muon spin relaxation: Introducing a new
  family of noncentrosymmetric superconductor that breaks time-reversal
  symmetry}.
\newblock \emph{\bibinfo{journal}{Phys. Rev. Lett.}}
  \textbf{\bibinfo{volume}{115}}, \bibinfo{pages}{267001}
  (\bibinfo{year}{2015}).

\bibitem{Singh2014}
\bibinfo{author}{Singh, R.~P.} \emph{et~al.}
\newblock \bibinfo{title}{Detection of time-reversal symmetry breaking in the
  noncentrosymmetric superconductor {Re}$_{6}${Zr} using muon-spin
  spectroscopy}.
\newblock \emph{\bibinfo{journal}{Phys. Rev. Lett.}}
  \textbf{\bibinfo{volume}{112}}, \bibinfo{pages}{107002}
  (\bibinfo{year}{2014}).

\bibitem{Shang2018b}
\bibinfo{author}{Shang, T.} \emph{et~al.}
\newblock \bibinfo{title}{Time-reversal symmetry breaking in {Re}-based
  superconductors}.
\newblock \emph{\bibinfo{journal}{Phys. Rev. Lett.}}
  \textbf{\bibinfo{volume}{121}}, \bibinfo{pages}{257002}
  (\bibinfo{year}{2018}).

\bibitem{Shang2020a}
\bibinfo{author}{Shang, T.} \emph{et~al.}
\newblock \bibinfo{title}{Time-reversal symmetry breaking in the
  noncentrosymmetric {Zr$_3$Ir} superconductor}.
\newblock \emph{\bibinfo{journal}{Phys. Rev. B}}
  \textbf{\bibinfo{volume}{102}}, \bibinfo{pages}{020503(R)}
  (\bibinfo{year}{2020}).

\bibitem{smidman2017}
\bibinfo{author}{Smidman, M.}, \bibinfo{author}{Salamon, M.~B.},
  \bibinfo{author}{Yuan, H.~Q.} \& \bibinfo{author}{Agterberg, D.~F.}
\newblock \bibinfo{title}{Superconductivity and spin--orbit coupling in
  non-centrosymmetric materials: {A} review}.
\newblock \emph{\bibinfo{journal}{Rep. Prog. Phys.}}
  \textbf{\bibinfo{volume}{80}}, \bibinfo{pages}{036501}
  (\bibinfo{year}{2017}).
\newblock \bibinfo{note}{And references therein}.

\bibitem{Bauer2012}
\bibinfo{editor}{Bauer, E.} \& \bibinfo{editor}{Sigrist, M.} (eds.)
  \emph{\bibinfo{title}{Non-Centrosymmetric Superconductors}}, vol.
  \bibinfo{volume}{847} (\bibinfo{publisher}{Springer Verlag},
  \bibinfo{address}{Berlin}, \bibinfo{year}{2012}).

\bibitem{Sato2017}
\bibinfo{author}{Sato, M.} \& \bibinfo{author}{Ando, Y.}
\newblock \bibinfo{title}{Topological superconductors: {A} review}.
\newblock \emph{\bibinfo{journal}{Rep. Prog. Phys.}}
  \textbf{\bibinfo{volume}{80}}, \bibinfo{pages}{076501}
  (\bibinfo{year}{2017}).

\bibitem{Qi2011}
\bibinfo{author}{Qi, X.-L.} \& \bibinfo{author}{Zhang, S.-C.}
\newblock \bibinfo{title}{Topological insulators and superconductors}.
\newblock \emph{\bibinfo{journal}{Rev. Mod. Phys.}}
  \textbf{\bibinfo{volume}{83}}, \bibinfo{pages}{1057--1110}
  (\bibinfo{year}{2011}).

\bibitem{Kallin2016}
\bibinfo{author}{Kallin, C.} \& \bibinfo{author}{Berlinsky, J.}
\newblock \bibinfo{title}{Chiral superconductors}.
\newblock \emph{\bibinfo{journal}{Rep. Prog. Phys.}}
  \textbf{\bibinfo{volume}{79}}, \bibinfo{pages}{054502}
  (\bibinfo{year}{2016}).

\bibitem{Sajilesh2020}
\bibinfo{author}{Sajilesh, K.~P.}, \bibinfo{author}{Singh, D.},
  \bibinfo{author}{Hillier, A.~D.} \& \bibinfo{author}{Singh, R.~P.}
\newblock \bibinfo{title}{Probing nodeless superconductivity in {La}${M}${Si}
  (${M}$ = {Ni}, {Pt}) using muon-spin rotation and relaxation}.
\newblock \emph{\bibinfo{journal}{Phys. Rev. B}}
  \textbf{\bibinfo{volume}{102}}, \bibinfo{pages}{094515}
  (\bibinfo{year}{2020}).

\bibitem{Sajilesh2018}
\bibinfo{author}{Sajilesh, K.~P.}, \bibinfo{author}{Singh, D.},
  \bibinfo{author}{Biswas, P.~K.}, \bibinfo{author}{Hillier, A.~D.} \&
  \bibinfo{author}{Singh, R.~P.}
\newblock \bibinfo{title}{Superconducting properties of the noncentrosymmetric
  superconductor {LaPtGe}}.
\newblock \emph{\bibinfo{journal}{Phys. Rev. B}} \textbf{\bibinfo{volume}{98}},
  \bibinfo{pages}{214505} (\bibinfo{year}{2018}).

\bibitem{Luke1993}
\bibinfo{author}{Luke, G.~M.} \emph{et~al.}
\newblock \bibinfo{title}{Muon spin relaxation in {${\mathrm{UPt}}_{3}$}}.
\newblock \emph{\bibinfo{journal}{Phys. Rev. Lett.}}
  \textbf{\bibinfo{volume}{71}}, \bibinfo{pages}{1466--1469}
  (\bibinfo{year}{1993}).

\bibitem{Dalmas1995}
\bibinfo{author}{de~R\'{e}otier, P.~D.} \emph{et~al.}
\newblock \bibinfo{title}{Absence of zero field muon spin relaxation induced by
  superconductivity in the {B} phase of {UPt3}}.
\newblock \emph{\bibinfo{journal}{Phys. Lett. A}}
  \textbf{\bibinfo{volume}{205}}, \bibinfo{pages}{239--243}
  (\bibinfo{year}{1995}).

\bibitem{Schemm2014}
\bibinfo{author}{Schemm, E.~R.}, \bibinfo{author}{Gannon, W.~J.},
  \bibinfo{author}{Wishne, C.~M.}, \bibinfo{author}{Halperin, W.~P.} \&
  \bibinfo{author}{Kapitulnik, A.}
\newblock \bibinfo{title}{Observation of broken time-reversal symmetry in the
  heavy-fermion superconductor {UPt}$_3$}.
\newblock \emph{\bibinfo{journal}{Science}} \textbf{\bibinfo{volume}{345}},
  \bibinfo{pages}{190--193} (\bibinfo{year}{2014}).

\bibitem{Suter2012}
\bibinfo{author}{A.~Suter, A.} \& \bibinfo{author}{Wojek, B.~M.}
\newblock \bibinfo{title}{Musrfit: {A} free platform-independent framework for
  $\mu${SR} data analysis}.
\newblock \emph{\bibinfo{journal}{Phys. Procedia}}
  \textbf{\bibinfo{volume}{30}}, \bibinfo{pages}{69} (\bibinfo{year}{2012}).

\bibitem{Zhu2008}
\bibinfo{author}{Zhu, X.}, \bibinfo{author}{Yang, H.}, \bibinfo{author}{Fang,
  L.}, \bibinfo{author}{Mu, G.} \& \bibinfo{author}{Wen, H.-H.}
\newblock \bibinfo{title}{Upper critical field, {H}all effect and
  magnetoresistance in the iron-based layered superconductor
  {LaFeAsO}$_{0.9}${F}$_{0.1-\delta}$}.
\newblock \emph{\bibinfo{journal}{Supercond. Sci. Technol.}}
  \textbf{\bibinfo{volume}{21}}, \bibinfo{pages}{105001}
  (\bibinfo{year}{2008}).

\bibitem{Werthamer1966}
\bibinfo{author}{Werthamer, N.~R.}, \bibinfo{author}{Helfand, E.} \&
  \bibinfo{author}{Hohenberg, P.~C.}
\newblock \bibinfo{title}{Temperature and purity dependence of the
  superconducting critical field, ${{H}}_{c2}$. {III}. {E}lectron spin and
  spin-orbit effects}.
\newblock \emph{\bibinfo{journal}{Phys. Rev.}} \textbf{\bibinfo{volume}{147}},
  \bibinfo{pages}{295} (\bibinfo{year}{1966}).

\bibitem{Gurevich2011}
\bibinfo{author}{Gurevich, A.}
\newblock \bibinfo{title}{Iron-based superconductors at high magnetic fields}.
\newblock \emph{\bibinfo{journal}{Rep. Prog. Phys}}
  \textbf{\bibinfo{volume}{74}}, \bibinfo{pages}{124501}
  (\bibinfo{year}{2011}).
\newblock \bibinfo{note}{And references therein}.

\bibitem{Shang2019}
\bibinfo{author}{Shang, T.} \emph{et~al.}
\newblock \bibinfo{title}{Enhanced ${T}_c$ and multiband superconductivity in
  the fully-gapped {ReBe}$_{22}$ superconductor}.
\newblock \emph{\bibinfo{journal}{New J. Phys.}} \textbf{\bibinfo{volume}{21}},
  \bibinfo{pages}{073034} (\bibinfo{year}{2019}).

\bibitem{Shang2020d}
\bibinfo{author}{Shang, T.} \emph{et~al.}
\newblock \bibinfo{title}{Multigap superconductivity in the {Mo$_5$PB$_2$}
  boron–phosphorus compound}.
\newblock \emph{\bibinfo{journal}{New J. Phys.}} \textbf{\bibinfo{volume}{22}},
  \bibinfo{pages}{093016} (\bibinfo{year}{2020}).

\bibitem{Amato1997}
\bibinfo{author}{Amato, A.}
\newblock \bibinfo{title}{Heavy-fermion systems studied by $\mu${SR}
  technique}.
\newblock \emph{\bibinfo{journal}{Rev. Mod. Phys.}}
  \textbf{\bibinfo{volume}{69}}, \bibinfo{pages}{1119--1180}
  (\bibinfo{year}{1997}).

\bibitem{Luke1998}
\bibinfo{author}{Luke, G.~M.} \emph{et~al.}
\newblock \bibinfo{title}{Time-reversal symmetry-breaking superconductivity in
  {Sr}$_{2}${RuO}$_{4}$}.
\newblock \emph{\bibinfo{journal}{Nature}} \textbf{\bibinfo{volume}{394}},
  \bibinfo{pages}{558} (\bibinfo{year}{1998}).

\bibitem{Aoki2003}
\bibinfo{author}{Aoki, Y.} \emph{et~al.}
\newblock \bibinfo{title}{Time-reversal symmetry-breaking superconductivity in
  heavy-fermion {PrOs}$_{4}${Sb}$_{12}$ detected by muon-spin relaxation}.
\newblock \emph{\bibinfo{journal}{Phys. Rev. Lett.}}
  \textbf{\bibinfo{volume}{91}}, \bibinfo{pages}{067003}
  (\bibinfo{year}{2003}).

\bibitem{Shang2018a}
\bibinfo{author}{Shang, T.} \emph{et~al.}
\newblock \bibinfo{title}{Nodeless superconductivity and time-reversal symmetry
  breaking in the noncentrosymmetric superconductor {Re}$_{24}${Ti}$_{5}$}.
\newblock \emph{\bibinfo{journal}{Phys. Rev. B}} \textbf{\bibinfo{volume}{97}},
  \bibinfo{pages}{020502} (\bibinfo{year}{2018}).

\bibitem{Shang2020b}
\bibinfo{author}{Shang, T.} \emph{et~al.}
\newblock \bibinfo{title}{Simultaneous nodal superconductivity and
  time-reversal symmetry breaking in the noncentrosymmetric superconductor
  {CaPtAs}}.
\newblock \emph{\bibinfo{journal}{Phys. Rev. Lett.}}
  \textbf{\bibinfo{volume}{124}}, \bibinfo{pages}{207001}
  (\bibinfo{year}{2020}).

\bibitem{Kubo1967}
\bibinfo{author}{Kubo, R.} \& \bibinfo{author}{Toyabe, T.}
\newblock \emph{\bibinfo{title}{Magnetic Resonance and Relaxation}}
  (\bibinfo{publisher}{North-Holland}, \bibinfo{address}{Amsterdam},
  \bibinfo{year}{1967}).

\bibitem{Yaouanc2011}
\bibinfo{author}{Yaouanc, A.} \& \bibinfo{author}{de~R\'eotier, P.~D.}
\newblock \emph{\bibinfo{title}{Muon Spin Rotation, Relaxation, and Resonance:
  Applications to Condensed Matter}} (\bibinfo{publisher}{Oxford University
  Press}, \bibinfo{address}{Oxford}, \bibinfo{year}{2011}).

\bibitem{TDOnote}
\bibinfo{note}{The preliminary measurements of magnetic penetration depth using
  the Tunnel Diode Oscillator method also indicate fully-gapped superconducting
  state consistent with the TF-muSR results and will be the subject of a
  separate article.}

\bibitem{sumita2018}
\bibinfo{author}{Sumita, S.} \& \bibinfo{author}{Yanase, Y.}
\newblock \bibinfo{title}{Unconventional superconducting gap structure
  protected by space group symmetry}.
\newblock \emph{\bibinfo{journal}{Phys. Rev. B}} \textbf{\bibinfo{volume}{97}},
  \bibinfo{pages}{134512} (\bibinfo{year}{2018}).

\bibitem{sumita2019}
\bibinfo{author}{Sumita, S.}, \bibinfo{author}{Nomoto, T.},
  \bibinfo{author}{Shiozaki, K.} \& \bibinfo{author}{Yanase, Y.}
\newblock \bibinfo{title}{Classification of topological crystalline
  superconducting nodes on high-symmetry lines: {P}oint nodes, line nodes, and
  {B}ogoliubov {F}ermi surfaces}.
\newblock \emph{\bibinfo{journal}{Phys. Rev. B}} \textbf{\bibinfo{volume}{99}},
  \bibinfo{pages}{134513} (\bibinfo{year}{2019}).

\bibitem{Ghosh2021}
\bibinfo{author}{Ghosh, S.~K.}, \bibinfo{author}{Annett, J.~F.} \&
  \bibinfo{author}{Quintanilla, J.}
\newblock \bibinfo{title}{Time-reversal symmetry breaking in superconductors
  through loop supercurrent order}.
\newblock \emph{\bibinfo{journal}{New J. Phys.}} \textbf{\bibinfo{volume}{23}},
  \bibinfo{pages}{083018} (\bibinfo{year}{2021}).

\bibitem{Weng2016}
\bibinfo{author}{Weng, Z.~F.} \emph{et~al.}
\newblock \bibinfo{title}{Two-gap superconductivity in {LaNiGa}$_2$ with
  nonunitary triplet pairing and even parity gap symmetry}.
\newblock \emph{\bibinfo{journal}{Phys. Rev. Lett.}}
  \textbf{\bibinfo{volume}{117}}, \bibinfo{pages}{027001}
  (\bibinfo{year}{2016}).

\bibitem{Ghosh2020b}
\bibinfo{author}{Ghosh, S.~K.} \emph{et~al.}
\newblock \bibinfo{title}{Quantitative theory of triplet pairing in the
  unconventional superconductor {${\mathrm{LaNiGa}}_{2}$}}.
\newblock \emph{\bibinfo{journal}{Phys. Rev. B}}
  \textbf{\bibinfo{volume}{101}}, \bibinfo{pages}{100506}
  (\bibinfo{year}{2020}).

\bibitem{Ghosh2022}
\bibinfo{author}{Ghosh, S.~K.}, \bibinfo{author}{Miyake, K.} \&
  \bibinfo{author}{Quintanilla, J.}
\newblock \bibinfo{title}{Spontaneous magnetization in a nonunitary triplet
  pairing state}.
\newblock \emph{\bibinfo{journal}{Unpublished}}  (\bibinfo{year}{2022}).

\bibitem{Shiroka2017}
\bibinfo{author}{Shiroka, T.} \emph{et~al.}
\newblock \bibinfo{title}{High-{$T_c$} superconductivity in undoped {ThFeAsN}}.
\newblock \emph{\bibinfo{journal}{Nat. Commun.}} \textbf{\bibinfo{volume}{8}},
  \bibinfo{pages}{156} (\bibinfo{year}{2017}).

\bibitem{Xia2006}
\bibinfo{author}{Xia, J.}, \bibinfo{author}{Maeno, Y.},
  \bibinfo{author}{Beyersdorf, P.~T.}, \bibinfo{author}{Fejer, M.~M.} \&
  \bibinfo{author}{Kapitulnik, A.}
\newblock \bibinfo{title}{{High Resolution Polar Kerr Effect Measurements of
  Sr$_2$RuO$_4$: Evidence for Broken Time-Reversal Symmetry in the
  Superconducting State}}.
\newblock \emph{\bibinfo{journal}{Phys. Rev. Lett.}}
  \textbf{\bibinfo{volume}{97}}, \bibinfo{pages}{167002}
  (\bibinfo{year}{2006}).

\bibitem{Uemura1991}
\bibinfo{author}{Uemura, Y.~J.} \emph{et~al.}
\newblock \bibinfo{title}{Basic similarities among cuprate, bismuthate,
  organic, {C}hevrel-phase, and heavy-fermion superconductors shown by
  penetration-depth measurements}.
\newblock \emph{\bibinfo{journal}{Phys. Rev. Lett.}}
  \textbf{\bibinfo{volume}{66}}, \bibinfo{pages}{2665} (\bibinfo{year}{1991}).

\bibitem{Domieracki2018}
\bibinfo{author}{Domieracki, K.} \& \bibinfo{author}{Kaczorowski, D.}
\newblock \bibinfo{title}{Superconductivity in non-centrosymmetric {ThNiSi}}.
\newblock \emph{\bibinfo{journal}{J. Alloys Compd.}}
  \textbf{\bibinfo{volume}{731}}, \bibinfo{pages}{64--69}
  (\bibinfo{year}{2018}).

\bibitem{Zhong1985}
\bibinfo{author}{Zhong, W.~X.}, \bibinfo{author}{Ng, W.~L.},
  \bibinfo{author}{Chevalier, B.}, \bibinfo{author}{Etourneau, J.} \&
  \bibinfo{author}{Hagenmuller, P.}
\newblock \bibinfo{title}{Structural and electrical properties of new
  silicides: {Th}{Co}$_x${Si}$_{2-x}$ (0 $\leq$ $x$ $\leq$ 1) and {Th}{T}{Si}
  ({T} = {Ni}, {Pt})}.
\newblock \emph{\bibinfo{journal}{Mater. Res. Bull.}}
  \textbf{\bibinfo{volume}{20}}, \bibinfo{pages}{1229--1238}
  (\bibinfo{year}{1985}).

\bibitem{Ptok2019}
\bibinfo{author}{Ptok, A.} \emph{et~al.}
\newblock \bibinfo{title}{Electronic and lattice properties of
  noncentrosymmetric superconductors {Th}${T}${Si} (${T}$ = {Co}, {Ir}, {Ni},
  and {Pt})}.
\newblock \emph{\bibinfo{journal}{Phys. Rev. B}}
  \textbf{\bibinfo{volume}{100}}, \bibinfo{pages}{165130}
  (\bibinfo{year}{2019}).

\bibitem{Xiao2021}
\bibinfo{author}{Xiao, G.} \emph{et~al.}
\newblock \bibinfo{title}{Superconductivity and strong spin-orbit coupling in a
  new noncentrosymmetric compound {ThIrP}}.
\newblock \emph{\bibinfo{journal}{Sci. China Phys. Mech. Astron.}}
  \textbf{\bibinfo{volume}{64}}, \bibinfo{pages}{107411}
  (\bibinfo{year}{2021}).

\bibitem{Liu2017}
\bibinfo{author}{Liu, J.} \& \bibinfo{author}{Balents, L.}
\newblock \bibinfo{title}{Correlation effects and quantum oscillations in
  topological nodal-loop semimetals}.
\newblock \emph{\bibinfo{journal}{Phys. Rev. B}} \textbf{\bibinfo{volume}{95}},
  \bibinfo{pages}{075426} (\bibinfo{year}{2017}).

\bibitem{Blundell1999}
\bibinfo{author}{Blundell, S.~J.}
\newblock \bibinfo{title}{Spin-polarized muons in condensed matter physics}.
\newblock \emph{\bibinfo{journal}{Contemp. Physics}}
  \textbf{\bibinfo{volume}{40}}, \bibinfo{pages}{175--192}
  (\bibinfo{year}{1999}).

\bibitem{Brandt2003}
\bibinfo{author}{Brandt, E.~H.}
\newblock \bibinfo{title}{Properties of the ideal {G}inzburg-{L}andau vortex
  lattice}.
\newblock \emph{\bibinfo{journal}{Phys. Rev. B}} \textbf{\bibinfo{volume}{68}},
  \bibinfo{pages}{054506} (\bibinfo{year}{2003}).

\bibitem{Tinkham1996}
\bibinfo{author}{Tinkham, M.}
\newblock \emph{\bibinfo{title}{Introduction to Superconductivity}}
  (\bibinfo{publisher}{Dover Publications}, \bibinfo{address}{Mineola, NY},
  \bibinfo{year}{1996}), \bibinfo{edition}{2} edn.

\bibitem{Prozorov2006}
\bibinfo{author}{Prozorov, R.} \& \bibinfo{author}{Giannetta, R.~W.}
\newblock \bibinfo{title}{Magnetic penetration depth in unconventional
  superconductors}.
\newblock \emph{\bibinfo{journal}{Supercond. Sci. Technol.}}
  \textbf{\bibinfo{volume}{19}}, \bibinfo{pages}{R41--R67}
  (\bibinfo{year}{2006}).

\bibitem{Carrington2003}
\bibinfo{author}{Carrington, A.} \& \bibinfo{author}{Manzano, F.}
\newblock \bibinfo{title}{Magnetic penetration depth of {Mg}{B}$_{2}$}.
\newblock \emph{\bibinfo{journal}{Physica C}} \textbf{\bibinfo{volume}{385}},
  \bibinfo{pages}{205--214} (\bibinfo{year}{2003}).

\end{thebibliography}

\end{document}